\documentclass{aa}  

\usepackage{graphicx}
\usepackage{txfonts}
\usepackage[dvipsnames]{xcolor}
\usepackage{subcaption}
\usepackage{multirow}
\usepackage[colorlinks=true,allcolors=blue]{hyperref}
\usepackage{soul}
\usepackage{array}

\newcommand{\oiii}{\hbox{[O\,{\sc iii}]}\xspace}
\newcommand{\Loiii}{\hbox{$L_{\rm [OIII]}$}\xspace}

\newcommand{\hb}{\hbox{H$\beta$}\xspace}

\newcommand{\ciii}{\hbox{C\,{\sc iii}]}\xspace}

\newcommand{\msun}{\ensuremath{\mathrm{M_\odot}}\xspace}	
\newcommand{\zsun}{\ensuremath{\mathrm{Z_\odot}}\xspace}
\newcommand{\lsun}{\ensuremath{\mathrm{L}_\odot}\xspace}	
\newcommand{\kms}{\ensuremath{\rm km~s^{-1}}\xspace}
\newcommand{\peryr}{\ensuremath{\rm yr^{-1}}\xspace}

\newcommand{\mstar}{\ensuremath{M_\star}\xspace}
	
\newcommand{\gs}{JADES-GS-z14-0\xspace}

\newcommand{\fesc}{\ensuremath{f_\mathrm{esc}}\xspace}
\newcommand{\prospector}{\texttt{prospector}\xspace}

\defcitealias{Tacchella:2022a}{T22}
\let\oldAA\AA % Allow \AA in math mode
\renewcommand{\AA}{\text{\oldAA}\xspace}
\newcommand{\prospmstar}{\textcolor{black}{\ensuremath{8.29^{+0.14}_{-0.09}}}\xspace}
\newcommand{\prosplogz}{\textcolor{black}{\ensuremath{-0.78^{+0.03}_{-0.03}}}\xspace}
\newcommand{\prospsfr}{\textcolor{black}{\ensuremath{1.16^{+0.05}_{-0.07}}}\xspace}

\newcommand{\prosptauv}{\textcolor{black}{\ensuremath{0.19^{+0.07}_{-0.05}}}\xspace}
\newcommand{\prospDLA}{\textcolor{black}{\ensuremath{21.96^{+0.08}_{-0.09}}}\xspace}

\begin{document}

   \title{The eventful life of a luminous galaxy at $z=14$: metal enrichment, feedback, and low gas fraction?}

   %\subtitle{I. Overviewing the $\kappa$-mechanism}

   \author{
Stefano Carniani \inst{1}
\and
Francesco D'Eugenio \inst{2,3,4}
\and
Xihan Ji \inst{2,3}
\and
Eleonora Parlanti \inst{1}
\and
Jan Scholtz \inst{2,3}
\and
Fengwu Sun \inst{5}
\and
Giacomo Venturi \inst{1}
\and
Tom J. L. C. Bakx \inst{6}
\and
Mirko Curti \inst{7}
\and
Roberto Maiolino \inst{2,3,8}
\and
Sandro Tacchella \inst{2,3}
\and
Jorge A. Zavala \inst{9}
\and
Kevin Hainline \inst{10}
\and
Joris Witstok \inst{11,3}
\and
Benjamin D.\ Johnson \inst{5}
\and
Stacey Alberts \inst{10}
\and
Andrew J.\ Bunker \inst{12}
\and
Stéphane Charlot \inst{13}
\and
Daniel J.\ Eisenstein \inst{5}
\and
Jakob M. Helton \inst{10}
\and
Peter Jakobsen \inst{11,14}
\and
Nimisha Kumari \inst{15}
\and
Brant Robertson \inst{16}
\and
Aayush Saxena \inst{12,8}
\and
Hannah \"Ubler \inst{2,3}
\and
Christina C. Williams \inst{17}
\and
Christopher N. A. Willmer \inst{10}
\and
Chris Willott \inst{18}
}

\institute{Scuola Normale Superiore, Piazza dei Cavalieri 7, I-56126 Pisa, Italy
 \and
Kavli Institute for Cosmology, University of Cambridge, Madingley Road, Cambridge, CB3 0HA, UK
 \and
Cavendish Laboratory, University of Cambridge, 19 JJ Thomson Avenue, Cambridge, CB3 0HE, UK
 \and
INAF -- Osservatorio Astronomico di Brera, via Brera 28, I-20121 Milano, Italy
 \and
Center for Astrophysics $|$ Harvard \& Smithsonian, 60 Garden St., Cambridge MA 02138 USA
 \and
Department of Space, Earth, \& Environment, Chalmers University of Technology, Chalmersplatsen 4 412 96 Gothenburg, Sweden
 \and
European Southern Observatory, Karl-Schwarzschild-Strasse 2, 85748 Garching, Germany
 \and
Department of Physics and Astronomy, University College London, Gower Street, London WC1E 6BT, UK
 \and
National Astronomical Observatory of Japan, 2 Chome-21-1, Osawa, Mitaka, Tokyo 181-8588, Japan
 \and
Steward Observatory, University of Arizona, 933 N. Cherry Avenue, Tucson, AZ 85721, USA
 \and
Cosmic Dawn Center (DAWN), Copenhagen, Denmark
 \and
Department of Physics, University of Oxford, Denys Wilkinson Building, Keble Road, Oxford OX1 3RH, UK
 \and
Sorbonne Universit\'e, CNRS, UMR 7095, Institut d'Astrophysique de Paris, 98 bis bd Arago, 75014 Paris, France
 \and
Niels Bohr Institute, University of Copenhagen, Jagtvej 128, DK-2200, Copenhagen, Denmark
 \and
AURA for European Space Agency, Space Telescope Science Institute, 3700 San Martin Drive. Baltimore, MD, 21210
 \and
Department of Astronomy and Astrophysics University of California, Santa Cruz, 1156 High Street, Santa Cruz CA 96054, USA
 \and
NSF’s National Optical-Infrared Astronomy Research Laboratory, 950 North Cherry Avenue, Tucson, AZ 85719, USA
 \and
NRC Herzberg, 5071 West Saanich Rd, Victoria, BC V9E 2E7, Canada
}

   %\date{Received September 15, 1996; accepted March 16, 1997}

% \abstract{}{}{}{}{} 
% 5 {} token are mandatory
 
  \abstract
  % {XXX}
  % {YYY}
  % {ZZZ}
  % {WWW}
  {JADES-GS-z14-0 is the most distant spectroscopically confirmed galaxy yet, at $z \gtrsim 14$. With a UV magnitude of --20.81, it is one of the most luminous galaxies at cosmic dawn and its half-light radius of 260 pc means that stars dominate the observed UV emission. We report the Atacama Large Millimeter/submillimeter Array (ALMA) detection of [O\,{\sc iii}]88$\mu$m line emission with a significance of 
  6.67$\sigma$ and at a frequency of 223.524~GHz, corresponding to a redshift of $14.1796\pm0.0007$, which is consistent with the candidate \ciii line detected in the NIRSpec spectrum. At this spectroscopic redshift, the Lyman-$\alpha$ break identified with NIRSpec requires a damped Lyman-$\alpha$ absorber with a column density of $\log(N_{\rm HI}/\mathrm{cm}^{-2})=21.96$.
  The total [O\,{\sc iii}]88$\mu$m luminosity ($\log(L_{\rm [OIII]}/\lsun) = 8.3\pm 0.1$) is fully consistent with the local $L_{\rm [OIII]}-SFR$ relation and indicating a gas-phase metallicity $>0.1~{\rm Z_{\rm \odot}}$. %Based on the ${L_{\rm [OIII]}/SFR}$ we infer a gas-phase metallicity $>0.1~{\rm Z_{\rm \odot}}$. \st{, which is somewhat unexpected given the weakness of the UV emission lines.} 
  Using \texttt{prospector} spectral energy distribution (SED) modeling and combining the ALMA data with JWST observations, we find  $Z=0.17~{\rm Z_{\rm \odot}}$ and a non-zero escape fraction of ionizing photons ($\sim11\%$), which is necessary by the code to reproduce the UV spectrum. We measure an ${\rm [O III] 5007\AA/[O III]88\mu{\rm m}}$ line flux ratio between 1 and 20, resulting in an upper limit to the electron density of roughly 700~cm$^{-3}$ assuming a single-cloud photoionization model. 
  %, which is lower than those measured in other high-$z$ luminous galaxies. 
  The \oiii emission line is spectrally resolved, with a FWHM of 102$^{+29}_{-22}\,$\kms, resulting in a dynamical mass of $\log$(M$_{\rm dyn}/\msun$) = 9.0$\pm0.2$. When compared to the stellar mass, this value represents a conservative upper limit on the gas mass fraction, which ranges from 50\% to 80\%, depending on the assumed star formation history. Past radiation-driven outflows may have cleared the galaxy from the gas, reducing the gas fraction and thus increasing the escape fraction of ionizing photons.}


   \maketitle
%
%-------------------------------------------------------------------

\section{Introduction}

One of the most exciting and puzzling coming out of observations with JWST is the discovery of numerous luminous ($M_{\rm UV}<-20$) galaxies already in place 300-500 Myr after the Big Bang \citep{Arrabal-Haro:2023a, Bunker:2023a, Curtis-Lake:2023, Robertson:2023, Wang:2023, Castellano:2024, Carniani:2024, Zavala:2024}. Indeed, the number of such bright galaxies at $z>10$ is up to one order of magnitude higher than what was extrapolated based on both pre-JWST observations and cosmological simulations \cite[e.g.,][]{Finkelstein:2023, Donnan:2024, Robertson:2024}. Such observations have raised several questions about the formation of first galaxy populations, and some studies have suggested a slow decline in the number density of galaxies at $z > 12$, with increasing efficiency of galaxy formation in halos at higher redshifts \citep[e.g.,][]{Dekel:2023, Robertson:2024}. Other studies suggest that the overabundance of ultraviolet-luminous galaxies can be also explained with an attenuation-free model \citep{Ferrara:2023, Ferrara:2023a}. This assumes that the radiation pressure of the stars expels gas and dust from the galaxy reducing the dust attenuation and boosting the observed UV emission. Alternatively, stochasticity in the galaxy luminosities and star-formation rate (SFR) at fixed halo mass \citep{Mason:2023, Mirocha:2023, Shen:2023}, a top-heavy initial mass function \citep{Inayoshi:2022, Trinca:2024}, and a contribution from accreting black holes \citep{Inayoshi:2022, Trinca:2024, Hegde:2024} are other possible scenarios that can explain the observed UV luminosity function at the Cosmic Dawn.

The study of the UV luminosity function alone is not sufficient to distinguish the various models. 
% We expect that every proposed model would leave a distinct imprint on the properties of the stellar population and interstellar medium of such luminous galaxies. 
%We expect that each proposed model would have different predictions on the properties of stellar populations and interstellar medium (ISM) of such luminous galaxies; constraining these quantities through observations would thus allow us to discern among the models.? 
However, each model is expected to make different predictions regarding the stellar population and interstellar medium (ISM) properties of such luminous galaxies; constraining these galaxies through observations would thus allow us to discern among the models and break the degeneracies encountered on the population level. 
However, the properties derived from the spectral energy distribution (SED) fitting of the most distant galaxies are not well-constrained and are highly degenerate between each other, particularly because the available photometry usually only probes the rest-frame UV emission (i.e., there is a strong degeneracy between the stellar mass, dust attenuation law, and star-formation history). To make further progress additional data at other wavelengths
are extremely valuable.

%the JWST observations of $z>10$ galaxies have revealed heterogeneous properties in terms of galaxy size and intensity of emission lines. Some galaxies are extremely compact ($<50$~pc), suggesting the presence of an accreting black hole, while others are more extended and their emission is associated with only the stellar population. NIRspec observations report strong rest-frame UV emission lines in GNz11 and GHz2, while no lines in \gs or other luminous galaxies. 
%

To date, JADES-GS-z14-0 is the most distant galaxy known so far \citep{Robertson:2024, Helton:2024, Carniani:2024}. With a UV luminosity of $M_{\rm UV}\sim-21$, this galaxy is among the most luminous galaxies at $z>8$ in GOODS-S and GOODS-N CANDELS fields \citep{Hainline:2023a}. It is two times more luminous than the GHZ2 at $z=12.33$ ($M_{\rm UV}\sim-20.5$; \citealt{Castellano:2024}) and nearly as luminous as GN-z11 at $z=10.6$ ($M_{\rm UV}\sim-21.5$; \citealt{Tacchella:2023,Bunker:2023a}).  
NIRCam observations reveal that JADES-GS-z14-0 is spatially resolved with a half-light radius of 260 pc (0.079$^{\prime\prime}$), indicating its emission is dominated by light from the stellar population and does not arise from an active galactic nucleus (AGN).  The galaxy has also been detected in a 43-h MIRI pointing at 7.7~$\mu$m, with excess wide-band emission likely indicating strong emission lines from \oiii$\lambda\lambda4959,5007$ and H$\beta$, corresponding to a rest-frame equivalent width $EW({\rm \oiii\lambda\lambda4959,5007+H\beta})= 370\AA$ \citep{Helton:2024}.

The JWST/NIRSpec follow-up observations have revealed a prominent sharp Lyman-$\alpha$ break, indicating a redshift of $z=14.32^{+0.08}_{-0.20}$ \citep{Carniani:2024}. 
As the Lyman-$\alpha$ break profile is sensitive to the column density of neutral hydrogen along the line of sight, the uncertainties on the spectroscopic redshift are quite large and asymmetric toward low values, whose probability is not negligible given the recent discovery of damped Lyman systems at very high redshifts \citep{Hsiao:2023, Heintz:2024,DEugenio:2023,Hainline:2024,Witstok:2024b}.  
Despite the deep observations, the NIRSpec spectrum does not show any clear rest-frame UV emission lines. There is only a tentative (3.6$\sigma$) detection of \ciii$\lambda \lambda 1907,1909$ at 2.89~$\mu$m, which would yield a redshift of 14.178 $\pm$ 0.013. If this redshift is confirmed, it would require the presence of a damped Lyman-$\alpha$ absorber (DLA) with a hydrogen column density of $\log(N_{\rm HI}/{\rm cm^{-2}})=22.23 $ to explain the redshifted Lyman-$\alpha$ break \citep{Carniani:2024}.

The SED fitting of JWST photometric and spectroscopic data indicates an ongoing star-formation rate (SFR) of $19~{\rm M_{\odot}/yr}$ and a stellar mass of $4\times10^{8}~{\rm M_{\odot}}$. The latter suffers from large uncertainties (0.4~dex) due to the unknown origin of the excess MIRI flux at 7 $\mu$m \citep{Helton:2024}. The spectral modelling suggests that one-third of the MIRI flux is associated with the rest-optical nebular emission lines \oiii$\lambda\lambda4959,5007$ and H$\beta$, but the results depend on the star-formation history priors adopted in the SED fitting \citetext{Which are highly uncertain at these high redshifts, \citealp{Tacchella:2022,Whitler:2023}}.

The absence of bright UV emission lines in NIRSpec spectrum sets only upper limits (with significant uncertainties) on the properties of the ISM. Additional multi-wavelength observations are thus fundamental to constrain the galaxy and ISM properties and provide deeper insights into the galaxy evolution of JADES-GS-z14-0.

In this paper, we combine JWST data with the public ALMA observations (DDT 2023.A.00037.S, PI: S. Schouws) of JADES-GS-z14-0. The observations carried out in the frequency range between 218 and 229 GHz allow us to identify the far-infrared oxygen line \oiii at rest-frame 88~$\mu$m and determine the redshift of the galaxy with high accuracy. In addition to that, the far-infrared line probes the gas-phase metallicity and then, combined with the MIRI data, allows us to constrain the electron density. Finally, the continuum emission allows us to study the dust content and investigate the nature of the moderate optical dust attenuation ($A_{\rm V}=0.3$) determined from the JWST data.

Throughout this paper, we adopt the standard cosmological parameters ${\rm H_0 = 70~km s^{-1}~Mpc^{-1}}$, ${\rm \Omega_{\rm M} = 0.30}$, ${\rm \Omega_{\rm \Lambda} = 0.70}$, according to which 1 arcsec at $z = 14$ corresponds to a proper distance of 3.268 kpc. Astronomical coordinates and magnitudes are given in the ICRS and AB systems, respectively.
Abundance patterns are from \cite{Byler:2017} unless otherwise stated.
%--------------------------------------------------------------------
\section{ALMA observations and data reduction}

\gs\ was observed with ALMA Band 6 as part of 2023.A.00037.S program, in two spectral configurations to target the \oiii\ 88 $\mu$m emission line and cover 98\% of the redshift probability distribution reported in \citet{Carniani:2024}. Each spectral configuration was observed for 2.82 hours of on-source integration time.
The calibrated visibilities were downloaded from the ALMA archive.
The first spectral configuration was observed with ALMA configuration C4, while the second one with configuration C5, hence the two spectral configurations have different spatial resolutions.

We performed the cleaning on the calibrated visibilities using the Common Astronomy Software Applications package (CASA; \citealt{Mcmullin:2007}) task \verb|tclean|. To optimize the sensitivity to detect the \oiii\ emission line, we adopted a natural and briggs robust=0.5 weighting scales, hogbom deconvolver, a spaxel scale of 0.1\arcsec, and a channel width of 10 \kms to create the final datacubes. The final datacubes for the first spectral configuration have a beam size of  0.82\arcsec\ $\times$ 0.60\arcsec\ and  1.09\arcsec\ $\times$ 0.82\arcsec\ for the briggs robust=0.5 and natural cubes, respectively. For the second spectrum configuration, we obtain datacubes with 0.42\arcsec $\times$ 0.38\arcsec\ and 0.58 \arcsec $\times$ 0.50 \arcsec\ for the briggs robust=0.5 and natural cubes, respectively.

We also imaged the dust-continuum emission at 88 $\mu$m rest-frame by using the ``mfs'' mode of the task tclean using natural weighting to maximize the S/N. The final continuum image has a beam size of 0.66\arcsec\ $\times$ 0.84\arcsec\ and a sensitivity of 4.4 $\mu$Jy~beam$^{-1}$.

\section{Data analysis}

%----------------------------------------------------------------- 
   \begin{figure*}[h!]
   \centering
   \includegraphics[width=1\linewidth]{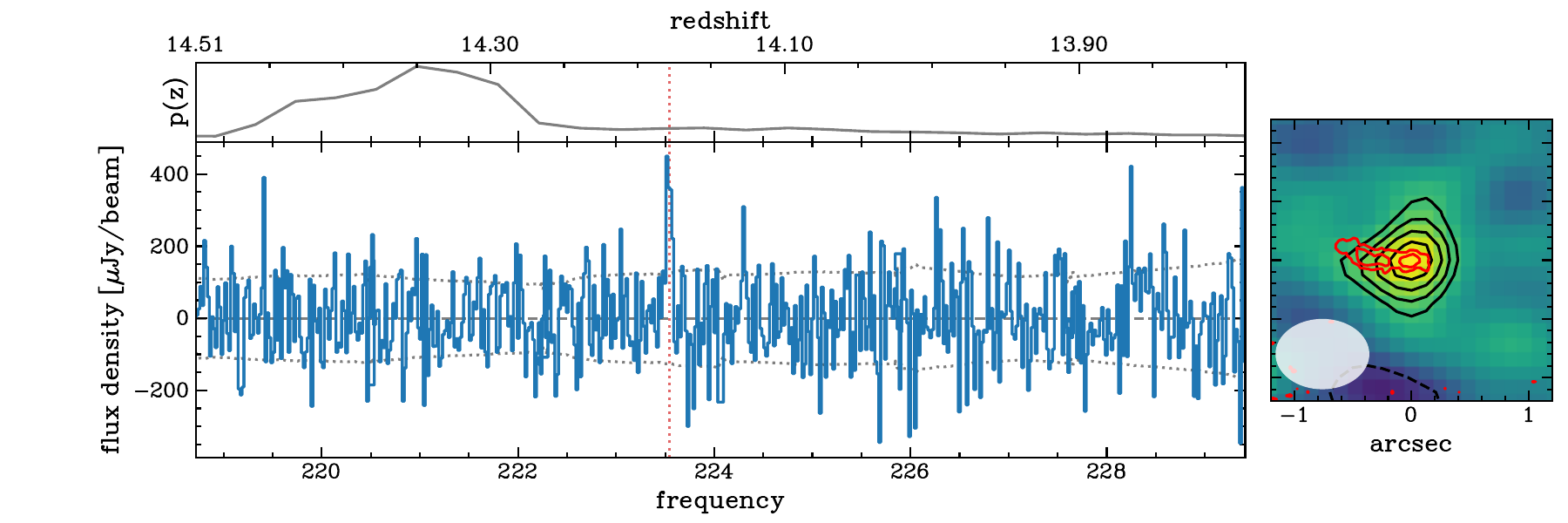}
      \caption{\oiii 88 $\mu$m spectrum (left) and flux map (right) of  \gs. The top panel illustrates the redshift probability distribution determined from the JWST/NIRSpec data. The red vertical dotted line shows the redshift determined from the candidate \ciii detection \citep{Carniani:2024}. The black contours in the flux map show  $\pm3$, $\pm4$, $\pm5$, and $\pm6$ $\sigma$ of \oiii emission, while red contours correspond to the rest-frame UV emission from JWST/NIRCam F200W observations.}
         \label{fig:spec}
   \end{figure*}

The frequency range of these ALMA observations covers about 98\% (Fig.~\ref{fig:spec}) of the posterior distribution function of the spectroscopic redshift determined from the fitting of the Lyman-break profile of JWST/NIRSpec data \citep{Carniani:2024}. Given the large uncertainty on the spectroscopic redshift, we have performed a blind-line search in the ALMA datacube. In particular, we have analyzed the region of the datacube enclosed within 5 arcsec from the location of the target. The adopted customized line finder code was already used in other ALMA studies \citep[e.g.,][]{Maiolino:2015, Carniani:2017, Carniani:2020} and was optimized to identify point-like sources in datacubes without defining a priori the line width of the emission line. More details of the code are explained in Appendix B of \cite{Carniani:2020}.  

At the location of the galaxy, we have identified an \oiii\ emission line with a significance level at 6.67$\sigma$ at a frequency of 223.524\ GHz, which corresponds to a redshift of  $14.1796\pm0.0007$. Fig.~\ref{fig:spec} shows the spectrum of the detected line and the integrated map. The line has a full width at half maximum (FWHM) of $102^{+29}_{-22}$~\kms and an integrated flux density of $0.037\pm 0.009$ Jy km s$^{-1}$, resulting in an \oiii\ luminosity log($\Loiii/L_\odot)=8.3\pm 0.1$.
We checked whether the detection could be due to a line associated to the foreground galaxy at $z = 3.475$ located close to \gs\ to its East \citep{Carniani:2024}. We find no line species that would be redshifted to a frequency of 223.524\ GHz given the above redshift.

We have also analyzed the continuum map, but no signal at the spatial position of the \oiii 88$\mu$m line emission has been detected. We have thus derived a $3\sigma$ upper limit on the continuum emission for a point-like source of $S_{\rm 88\mu m}<13~{\rm \mu Jy}$.

The fidelity of the ALMA emission line detection is verified in Appendix~\ref{apd:stats}. 
We note that there is an additional 3.7$\sigma$ signal in the \oiii moment-0 map at the same redshift 2.5 arcsec NW from \gs. Given the low significance and the lack of continuum detection in the NIRCam images, we do not consider this additional 3.7-$\sigma$ signal as a solid detection.

\section{Results}
\subsection{Dust content}
As the dust thermal emission is not detected in the ALMA continuum map at 88$\mu$m, we estimate an upper limit on dust mass. We assume a typical single-temperature modified black body function (e.g., equation 1 in \citealt{Carniani:2019}) to reproduce the dust thermal emission spectral-energy distribution in the rest-frame far-infrared. We adopt a dust opacity coefficient at 250 GHz of 0.45 g cm$^{-2}$ and an emissivity index of $\beta=1.8$ \citep{Witstok:2023}, and correct for the effect of the cosmic microwave background following \cite{daCunha:2013}. Given the modified black body function, the flux at 88$\mu$m depends on dust mass and temperature.

\begin{figure}
   \centering
   \includegraphics[width=0.9\linewidth]{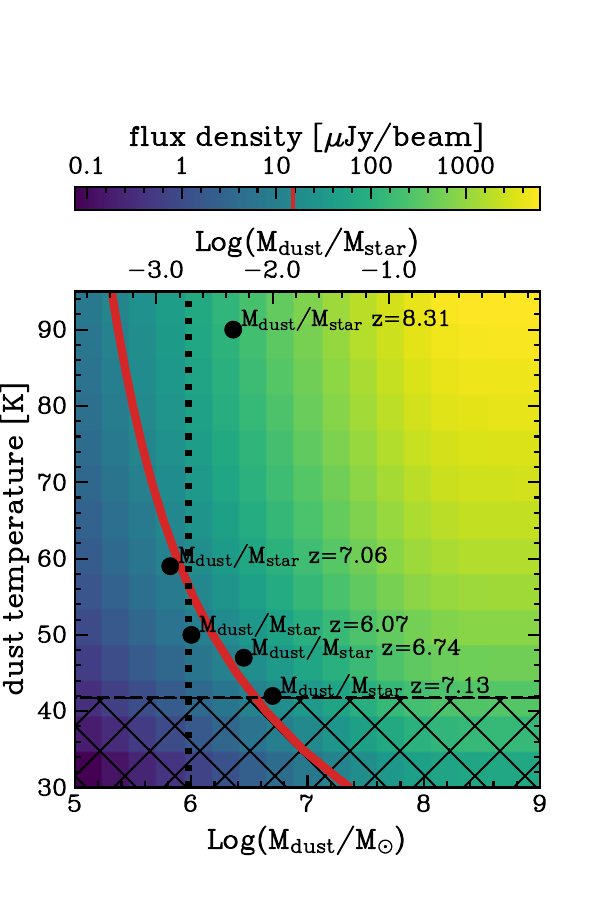}
      \caption{Predicted continuum flux density at the rest-frame wavelength 88$\mu$m as a function of dust temperature and dust mass. The red line shows the 3$\sigma$ sensitivity of the DDT ALMA program of \gs. The top axis shows the dust-to-stellar mass ratio for the stellar mass of \gs ($M_{\rm star} = 10^{8.6}~{\rm M_\odot}$). The vertical dashed line shows the expected dust-to-stellar mass ratio for the supernova dust enrichment mechanism, and black circles illustrate the current measurements in galaxies at  $z>7$. The CMB temperature at $z=14$ is $41.7$~K (dashed region).}
         \label{fig:dust}
   \end{figure}

Figure~\ref{fig:dust} shows the $3\sigma$ upper limits on the dust mass as a function of the dust temperature, based on the continuum sensitivity of the DDT ALMA program. The dust mass cannot be constrained with only one ALMA measurement due to the degeneracy with the dust temperature. In addition to that, if we assume that the dust temperature is lower than 60 K, the continuum sensitivity is not sufficient to put any constraint on the dust enrichment mechanism.  The upper limits on the dust-to-stellar mass ratio at low temperatures are higher than 0.002 (vertical black dashed line in Figure \ref{fig:dust}). Such value is higher than the typical dust-to-stellar mass ratio observed in high-$z$ galaxies \citep{Witstok:2022} and expected by theoretical models assuming that at such early times dust enrichment mechanism is mainly associated with supernova ejecta \citep{Schneider:2023}. This means that in the hypothesis of low temperatures, the depth of current observations is not sufficient to put any constraints on the dust content. 

We note that the upper limit on the continuum emission is consistent with the recent predictions by \cite{Ferrara:2024}. The authors show that the flux at 88~$\mu$m depends on both the dust-to-stellar mass ratio and the extension of the dust. In their fiducial case with dust-to-stellar equal to 1/529, the dust is expected to be extended up to 1.4 kpc to reproduce the inferred $A_{\rm V}$ from \cite{Carniani:2024}, have a luminosity-weighted temperature of 72~K, and a flux of about 15~$\mu$Jy.

In conclusion, the continuum sensitivity of this DDT program does not provide any information about the dust content in \gs and deeper future observations are needed to unveil the dust content in the galaxy and investigate the formation (and destruction) of dust grains in the early Universe.

%----------------------------------------------------------------- 
%-----------------------------------------------------------------

\subsection{DLA}\label{sec:dla}

We detect an emission line at 6.7$\sigma$ which we associate with the \oiii\,88\,$\mu$m\ transition, implying a spectroscopic redshift of $z = 14.1796$ $\pm$ 0.0007. The \oiii redshift is within the probability distribution function of redshifts inferred from the Lyman break analysis for this galaxy (top panel of Figure~\ref{fig:spec}) and is consistent with the redshift from the tentative detection of \ciii of 14.178 $\pm$ 0.013 \citep{Carniani:2024}. This redshift, however, is lower than what was predicted from the Ly$\alpha$ break alone ($z = 14.32$), and indicates the presence of additional damped Ly$\alpha$ absorption for the source, in agreement for what has been observed in other galaxies at $z > 10$ \citep{Heintz:2023, Umeda:2024, DEugenio:2024, Hainline:2024, Heintz:2024, Witstok:2024b}.

Using the redshift determined from the detected far-infrared line, we explore the potential need for a DLA to explain the Lyman break profile. To that end, we fit the slit-loss-corrected NIRSpec spectrum with a power-law function of the form $f_\lambda\propto\lambda^{-\beta}$ and apply the attenuation we would expect from the host galaxy ISM or circumgalactic medium column density of neutral hydrogen ($N_{\rm HI}$) along the line of sight as modeled in \cite{Witstok:2024b} and \cite{Hainline:2024}. From this fit, we measure a hydrogen column density of $\log(N_{\rm HI}/\mathrm{cm}^{-2})=22.23$ for JADES-GS-z14-0. %$$, which is in agreement with that determined from the SED fitting (see Section~\ref{sec:sed}). We adopt this value as the fiducial value.

This column density, which is similar to those measured in other, intrinsically fainter, high-redshift galaxies, indicates the presence of a large reservoir of pristine gas surrounding JADES-GS-z14-0. The presence of a strong DLA in JADES-GS-z14-0 may indicate that we are observing the source still in the process of assembly and growth. This result may be surprising given its observed brightness and the lack of similar sources found at later cosmic times \citep{Carniani:2024}. The location, and extent of the DLA with respect to the host galaxy is still uncertain, however. 
%In Section \ref{sec:escape}, we discuss the escape fraction of ionizing radiation in JADES-GS-z14-0 and the potential for a fast outflow in this source, which may have contributed to the DLA we observe. 

The smoothness of the Ly$\alpha$ break, implied by the redshift measured via \oiii and \ciii being slightly lower than expected from the break alone, can alternatively be explained by invoking an unusually strong two-photon nebular continuum \citep{2024arXiv240803189K}. This would require the nebular continuum to dominate in the rest-frame UV, which only occurs in the presence of very hot ionising sources ($T_\text{eff} \gtrsim 10^5 \, \mathrm{K}$) and at gas densities of $n \lesssim 10^5 \, \mathrm{cm^{-3}}$ \citep{Cameron:2023}. In this case, the two-photon radiative decay from the $2s$ state produces prominent continuum emission with a characteristic, fixed shape featuring a smooth `rollover' redwards of Ly$\alpha$ similar to DLA absorption \citep{1951ApJ...114..407S}. %Given its remarkably low gas fraction (Section~\ref{sec:Mdyn}) and potentially high ionising-photon escape fraction (Section~\ref{sec:escape}), both of which may argue against the large column densities of neutral hydrogen implied in the DLA scenario, the two-photon continuum may indeed play a role in shaping the Ly$\alpha$ break of JADES-GS-z14-0.

\subsection{[OIII] 88\texorpdfstring{$\,\mu$}{u}m luminosity}

%----------------------------------------------------------------- 
   \begin{figure}
   \centering
   \includegraphics[width=0.9\linewidth]{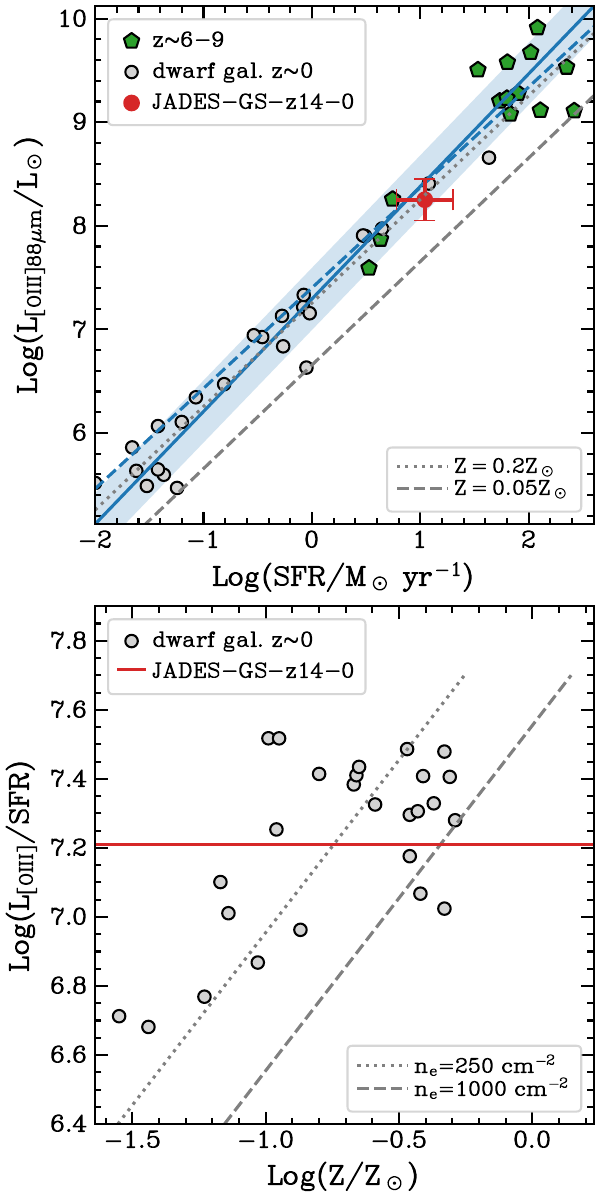}
      \caption{$L_{\rm [OIII]}$ versus SFR (top panel) and  $L_{\rm [OIII]}/\mathrm{SFR}$ versus gas-phase metallicity (bottom panel). The red circle and horizontal line (in top and bottom panels, respectively) mark the measurement for \gs. Gray circles are local dwarf galaxies, while green pentagons are $z\sim6-9$ galaxies. The solid blue line shows the local relation by \cite{DeLooze:2014} for metal-poor galaxies, with the shaded area corresponding to the 1$\sigma$ uncertainty. The blue dashed line illustrates the $L_{\rm [OIII]}-\mathrm{SFR}$  relation for high-$z$ galaxies by \cite{Harikane:2020}. Dotted and dashed gray lines illustrate the Cloudy models for different gas-phase metallicities and electron densities \citep{Harikane:2020, Jones:2020}.
              }
         \label{fig:oiii_SFR}
   \end{figure}
%-----------------------------------------------------------------

We inferred the \oiii\ 88\,$\mu$m luminosity from the 6.7$\sigma$ detection at 223.524 GHz. We estimated the measured \oiii\ luminosity by following \cite{Carilli:2013} and corrected for a modest lensing magnification factor of 1.17 \citep{Carniani:2024}, resulting in log$_{10}$(L$_{\rm [OIII]}$/L$_{\odot}$) = $8.3\pm0.1$. We compare the \oiii\ luminosity to the SFR in the top panel of Figure \ref{fig:oiii_SFR}, alongside other high-redshift detections \citep{Harikane:2020, Witstok:2022, Fujimoto:2024} and local low-metallicity dwarf galaxies \citep{DeLooze:2014}. 

The \oiii detection of \gs falls slightly below the local and high-$z$ \oiii-SFR observed relations but they agree within the uncertainties. This excludes that \gs is extremely metal-poor ($Z<0.05~Z_{\odot}$).
Indeed, as shown in the bottom panel of Figure \ref{fig:oiii_SFR}, the \oiii depends both on SFR and gas-phase metallicity, and galaxies with $Z<0.05~Z_{\odot}$ fall below the local relation. 
For example, IZw18 and SBS0335-052, which have a gas-phase metallicity of about 3\% solar \citep{DeLooze:2014}, show a  \Loiii deficit of 0.2-0.3 dex relative to the local relation. The ratio between \Loiii and SFR of \gs is similar to those measured in dwarf galaxies with $0.05~Z_{\odot}<Z<1~Z_{\odot}$. However, we cannot exclude a lower metallicity scenario given that \Loiii/SFR  is also sensitive to the ionization parameter and electron density. 

\begin{figure}
   \centering
   \includegraphics[width=\linewidth]{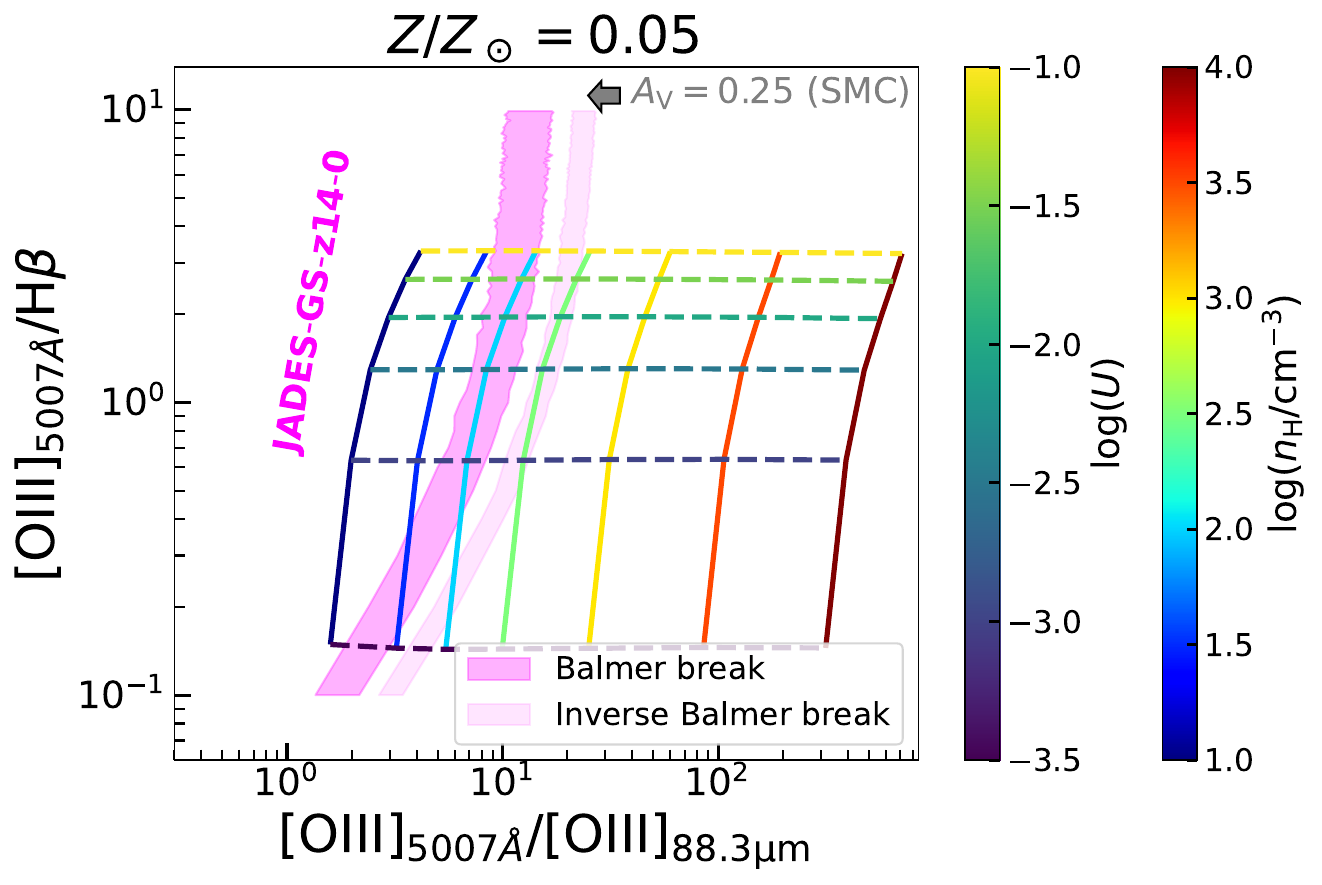}
   \includegraphics[width=\linewidth]{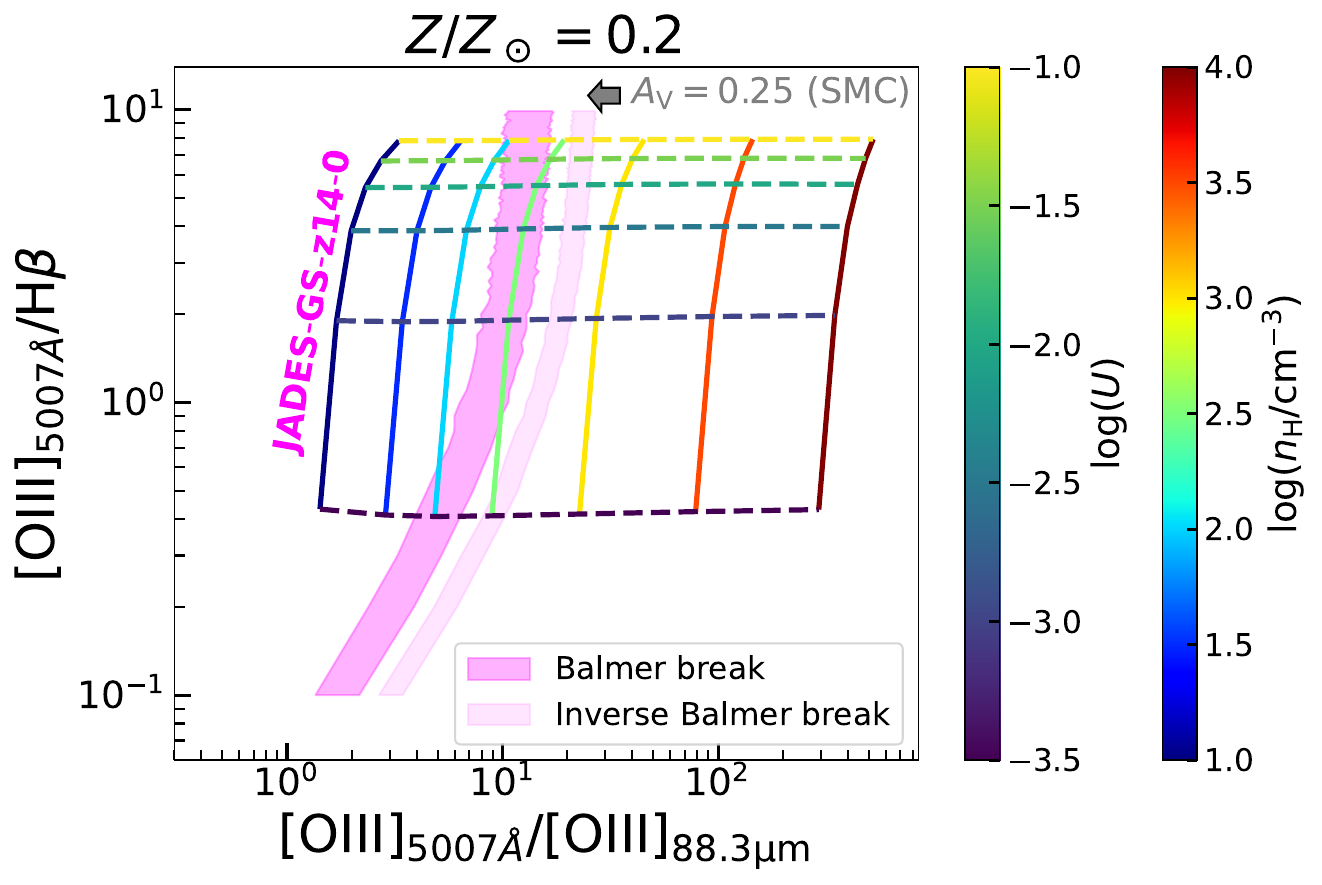}
   
      \caption{ \oiii$5007\AA$/\oiii$88\mu$m versus \oiii$5007\AA$/\hb diagram. Top and bottom panels show model grids computed with \textsc{Cloudy} with metallicities $Z/Z_\odot = 0.05$ to $Z/Z_\odot = 0.2$, respectively.
      Each model grid spans a range of ionization parameters ($-3.5\leq \log U \leq -1$) and hydrogen densities ($10~{\rm cm^{-3}}\leq n_{\rm H} \leq 10^4~{\rm cm^{-3}}$) in steps of 0.5 dex.
      Details on model parameters are provided in Tab.~\ref{tab:models}.
      The magenta-shaded regions show the plausible attenuation-corrected line ratios for JADES-GS-z14-0 inferred from observations by JWST/MIRI and ALMA. {The dark regions report the range of line ratios assuming that the MIRI excess is only due to the optical line. The light regions illustrate the range of line ratios in the case the stellar age is 1~Myr and there is an inverse Balmer break in the spectrum.}
      The gray arrow represents a reddening vector with $A_{\rm V} = 0.25$ assuming the nebular dust attenuation follows an average SMC extinction curve.
      %\xj{TBD: show a reddening vector or apply dust correction (or nothing)?}
      }
         \label{fig:o33o3hb}
   \end{figure}

In Figure \ref{fig:o33o3hb} we combine JWST/MIRI and ALMA data to determine the region covered by \gs\ in the $L_{\rm [OIII]5007\AA}/L_{\rm H\beta}$ versus $L_{\rm [OIII]5007\AA}/L_{\rm [OIII]88\mu m}$ diagram. 
Since \oiii$\lambda 5007$ and \hb are not spectrally resolved in MIRI, we vary $L_{\rm [OIII]5007\AA}/L_{\rm H\beta}$ from 0.1 to 10, which is a plausible range for metal-poor galaxies with different ionization conditions.
For each combination of $L_{\rm [OIII]5007\AA}/L_{\rm H\beta}$, we use the MIRI excess flux at 7~$\rm \mu m$ ($\Delta f = 27.5\pm 5.6$ nJy; \citealt{Helton:2024}) as the total flux of the blend, and calculate the possible value of $L_{\rm [OIII]5007\AA}$ and thus $L_{\rm [OIII]5007\AA}/L_{\rm [OIII]88\mu m}$. The dark magenta-shaded region reported in Figure~\ref{fig:o33o3hb} shows the range of line ratios assuming that the stellar age of the galaxy is $50-100 Myr$ and thus the stellar continuum emission 
 at 7.7~$\mu$m is as strong as that at 4.4~$\mu$m (see details in Appendix~\ref{app:balmer_break}). The light magenta-shaded regions illustrate the extreme case in which there is an inverse Balmer break in the spectral due to a young ($<5$~Myr) stellar population in the galaxy (Appendix~\ref{app:balmer_break}). We stress that the $L_{\rm [OIII]5007\AA}$ and $L_{\rm H\beta}$ inferred directly from the SED fitting cannot be used in this analysis because the fitting process and thus the intensity of the emission lines are inferred assuming a fixed electronic density.
 
The line ratios we show are corrected for dust attenuation. To illustrate the effect of dust attenuation, we plot a reddening vector corresponding to $A_{\rm V} = 0.25$ (as extreme value determined by the SED fitting later in Sec.~\ref{sec:sed}) assuming an average SMC extinction law \citep{gordon_2003smc}.
Due to the low $A_{\rm V}$, the reddening correction for $L_{\rm [OIII]5007\AA}/L_{\rm [OIII]88\mu m}$ is not sensitive to the choice of the extinction curve.

For comparison, we also show in Figure~\ref{fig:o33o3hb} predictions from photoionization single-density component models computed with the code \textsc{Cloudy} \citep[c17.03,][]{Ferland:2017} with a range of metallicities, ionization parameters, and gas densities.
We summarize the full set of model parameters in Tab.~\ref{tab:models}.
In brief, we assume the gas is photoionized by radiation from a 1 Myr-old simple stellar population (SSP) generated by the Binary Population and Spectral Synthesis code \citep[BPASS v2.3,][]{eldridge_bpass17,stanway_bpass18,byrne_bpass22}.
We allow the stellar metallicity to vary from 5\% solar to 20\% solar and match the gas-phase metallicity with the stellar metallicity.
Dust depletion is included such that 40\% of oxygen from the gas is depleted onto dust grains \citep{cowie_dust1986,jenkins_dust1987}.
We vary the dimensionless ionization parameter, defined as $U\equiv Q_0/n_{\rm H}c$ ($Q_0$ is the ionizing photon flux, $n_{\rm H}$ is the hydrogen density, and $c$ is the speed of light), from $-3.5$ to $-1$, and the density in a range of $10\leq n_{\rm H}/\rm cm^{-3} \leq 10^4$.
For the equation of state (EoS), we assume the gas is isobaric.

The constraints imposed by the ALMA and MIRI observations set upper limits on the electron density ($n_{\rm e}$) of [O\,{\sc iii}]88$\mu$m-emitting regions of \gs\ depending on the metallicity.
By interpolating the model grids linearly in the logarithmic space of $n_{\rm H}$ and $U$ using the \textsc{griddata} function in the \textsc{scipy} package \citep{2020SciPy-NMeth}, we obtain 
density upper limits of $n_{\rm H; uplim} = 180~{\rm cm^{-3}}$ ($Z/Z_\odot = 0.05$), $300~{\rm cm^{-3}}$ ($Z/Z_\odot = 0.1$), and $420~{\rm cm^{-3}}$ ($Z/Z_\odot = 0.2$) after dust attenuation corrections. 
In the inverse Balmer break scenario, which corresponds to a stellar age of less than 5\,Myr, we determine an upper limit for the electronic density of \(n_H^{\text{uplim}} = 700 \, \text{cm}^{-3}\) at \(Z/Z_{\odot} = 0.2\). We emphasize that this upper limit is highly conservative, even when considering the possibility of multiple-density regions within the galaxy. Specifically, if a fraction of the [O\,\textsc{iii}]\,5007 emission arises from regions with electronic densities exceeding the critical density of [O\,\textsc{iii}]\,88\,\(\mu\)m (\(\sim 500\text{--}1000 \, \text{cm}^{-3}\)), where [O\,\textsc{iii}]\,88\,\(\mu\)m is not emitted, the \(L_{\text{[O\,\textsc{iii}]}\,5007\,\text{\AA}}/L_{\text{[O\,\textsc{iii}]}\,88\,\mu\text{m}}\) ratio and consequently the inferred electronic density would be lower than the value reported in Figure~\ref{fig:o33o3hb}. In summary, we conclude that the [O\,\textsc{iii}]\,88\,\(\mu\)m-emitting regions in JADES-GS-z14-0 have an electronic density below \(700 \, \text{cm}^{-3}\).  However, \gs may also contain denser star-forming regions that are not effectively probed by [O\,\textsc{iii}]\,88\,\(\mu\)m.  High-spectral-resolution observations of  C\,\textsc{iii}] doublet could provide a valuable diagnostic for the electronic densities in these regions \citep{James:2014, Mingozzi:2022}.

The upper limits on the densities determined from the optical [O\,\textsc{ii}]$\lambda\lambda$3726,3729 doublet, which has a critical density comparable to that of [O\,\textsc{iii}]88$\mu$m, are comparable to those found in galaxies at $z = 2-3$ \citep[e.g.,][]{sanders_2016ne,kashino_2017ne,strom_2017ne,davies_2021ne},
but appear lower than galaxies at $z > 8$ \citep[e.g.,][]{isobe_nez2023,abdurrouf_2024ne, Marconcini+2024}. For example, \citet{isobe_nez2023} find that galaxies at $z=4-9$ have electron densities higher than $n_{\rm e} \sim 300~{\rm cm^{-3}}$ ($n_{\rm e} \approx 1.1~n_{\rm H}$ for H\,{\sc ii} regions), and the median value for $z=7-9$ is $n_{\rm e} \sim 1000~{\rm cm^{-3}}$. Furthermore, $n_{\rm e}$ might further increase with increasing $z$ if the relation given by \citet{isobe_nez2023} is extrapolated to $z>9$.
As another example, based on the ALMA observation of GHZ2, a bright galaxy at $z=12.33$, \citet{Zavala:2024} infer a density of $n_{\rm e}=100-3000~{\rm cm^{-3}}$ or higher depending on the assumed temperature.
However, we caution that the above high-redshift results are currently based on small number statistics and might be subject to measurement systematics \citep{sanders_2024ne}.

%If we exclude the region of the diagram with $n_{\rm e}$, ALMA data suggest an gas-phase metallicity higher than 20\% solar that would be in contrast with the lack of UV emission line in the deep NIRSpec spectrum if we do not assume high escape fraction. 
From Figure~\ref{fig:o33o3hb}, one can see a degeneracy between the ionization parameter and the metallicity.
If we constrain $n_{\rm e} \gtrsim 300~{\rm cm^{-3}}$ \citep[i.e., to be higher than the average density in SF galaxies at $z=2-3$,][]{sanders_2016ne,strom_2017ne}, then only the highest metallicity model with $Z/Z_\odot = 0.2$ [$\rm (O/H)_{\rm gas}/(O/H)_{\rm solar} = 0.12$ after dust depletion] is able to reproduce the observed line ratios, which also constrain the ionization parameter to be $-3 \lesssim \log U \lesssim -1.5$.
Considering an observed scatter of roughly $100~{\rm cm^{-3}}$ in the density \citep{sanders_2016ne,strom_2017ne}, models with $Z/Z_\odot = 0.1$ [$\rm (O/H)_{\rm gas}/(O/H)_{\rm solar} = 0.06$] and $-3 \lesssim \log U \lesssim -1.5$ are also plausible.
Again, we caution that the current constraint on the densities at high redshift has large uncertainties and we cannot exclude the possibility that JADES-GS-z14-0 is a low-density outlier at high redshift.
%\xj{TODO: update with dust attenuation correction}

%----------------------------------------------------------------- 

\begin{table}
        \centering
        \caption{Input parameters for \textsc{Cloudy} photoionization models}
        \label{tab:models}
        \begin{tabular}{l c}
            \hline
            \hline
            Parameter & Values \\
            \hline
            $Z/Z_\odot $ & 0.05, 0.1, 0.2\\
            \hline
            $\log U$& $-3.5$, $-3$, $-2.5$, $-2$, $-1.5$, $-1$ \\
            \hline
            $\log (n_{\rm H}/{\rm cm^{-3}})$& 1, 1.5, 2, 2.5, 3, 3.5, 4 \\
            \hline
            EoS & Constant gas pressure \\
            \hline
            SED & 1 Myr SSP from BPASS \\
            \hline
            Solar abundance set & \citet{grevesse_solar10} abundance set\\
            \hline
            N/O (C/O) vs. O/H & \citet{groves_no04} relation\\
            \hline
            He/H vs. O/H & \citet{dopita_heh00} relation\\
            \hline
            Dust depletion & \citet{cowie_dust1986};\\
             & \citet{jenkins_dust1987} depletion factors\\
            \hline
            Atomic data& CHIANTI (v7; \citealp{chianti_old},\\
            & \citealp{chianti_v7})\\
            \hline
        \end{tabular}
\end{table}

%----------------------------------------------------------------- 

%----------------------------------------------------------------- 

%-----------------------------------------------------------------

%-----------------------------------------------------------------
\section{SED modelling: NIRSpec+NIRCam+MIRI+ALMA}\label{sec:sed}
\subsection{Stellar mass}
\begin{figure*}
    \includegraphics[trim={10cm 0 10cm 0},clip,width=\textwidth]{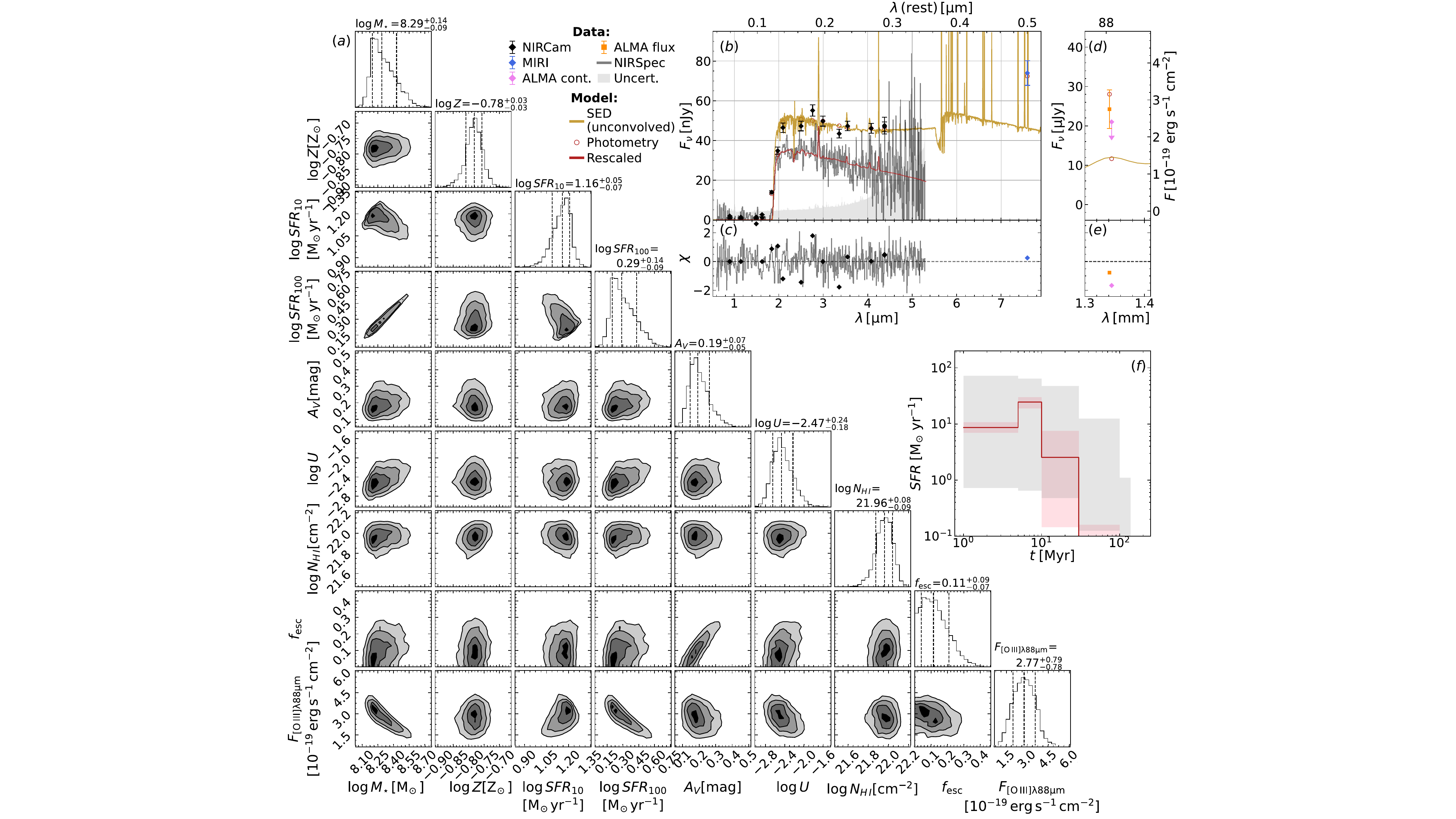}
  {\phantomsubcaption\label{fig:sed.a}
   \phantomsubcaption\label{fig:sed.b}
   \phantomsubcaption\label{fig:sed.c}
   \phantomsubcaption\label{fig:sed.d}
   \phantomsubcaption\label{fig:sed.e}
   \phantomsubcaption\label{fig:sed.f}}
  \caption{Summary of \prospector SED model. Panel~\subref{fig:sed.a}. Triangle diagram with the marginalised posterior distribution over a subset of the model-free or dependent parameters (see Table~\ref{table:2} for a description of the parameters and their probability prior).
  Panel~\subref{fig:sed.b}. SED, including wide- and medium-band measured flux densities (diamonds), the spectrum (grey line with grey region as the uncertainty). NIRCam, MIRI and ALMA data are in black, blue and pink, respectively.
  Model predictions are red circles (flux densities) or the red line (spectrum). The sand-coloured line is the model SED without convolving to the spectral resolution fo the data.
  Panel~\subref{fig:sed.c}. Model residuals normalised by the uncertainties.
  Panel~\subref{fig:sed.d}. Same as panel~\subref{fig:sed.b}, but for the FIR region. The orange square is the ALMA \oiii flux.
  Panel~\subref{fig:sed.e}. Same as panel~\subref{fig:sed.c}, but for the FIR region. 
  Panel~\subref{fig:sed.f}. Star-formation history (red), with the prior probability in grey.
  }\label{fig:sed}
\end{figure*}

The combination of ALMA and JWST provides us with new constraints on the SED modelling. We add our ALMA constraints to existing NIR and MIR data; we use NIRSpec/MSA spectroscopy and NIRCam photometry from \cite{Carniani:2024}, in combination with MIRI F770W photometry from \citet[][capturing \hb and \oiii$\lambda \lambda 4959,5007$]{Helton:2024}. We adopt the Bayesian inference framework \texttt{prospector} \citep{Johnson:2021}, itself relying on \texttt{fsps} \citep{Conroy2009, Conroy2010} for stellar spectra, and on pre-computed nebular emission from \citet{Byler2017}. In our setup, we use MIST isochrones \citep{choi+2016} and the C3K model atmospheres \citep{conroy+2019}.
We use a recent implementation\footnote{In particular, we employed development branch `v2', build 1.2.1.dev103+gb055d6f.d20240919.}. We generally follow the model setup of \citet[][hereafter \citetalias{Tacchella:2022a}]{Tacchella:2022a}. The differences are the IMF \citetext{we assume a \citealp{Chabrier:2003} IMF with lower and upper mass limits of 0.1 and 300~\msun}, a stronger prior on the parameters of the \citet{Kriek:2013} dust attenuation, the inclusion of a 2\textsuperscript{nd}-order multiplicative Chebyshev polynomial (to match the shapes of the photometry and of the spectrum), the maximum allowed redshift (no star formation is allowed earlier than $z=30$), the addition of a DLA (with variable redshift and column density), and a rising prior on the star-formation history (SFH), following the redshift evolution of the accretion rate of dark-matter haloes. The rising prior is parameterized by the log-ratio of the SFR between each time bin and the previous bin in cosmic time; by default, this prior behaves similarly to the delayed-$\tau$ models, but we allow substantial freedom by using a probability prior on the log SFR ratio that is a Student's $t$ distribution with scale 1 and $\nu=2$ \citep[following][]{Leja:2019}. 
The rising prior on the SFH follows the mass accretion rate of dark-matter halos \citep{Tacchella:2018}, and will be presented in \cite{Turner:2024}. The shape of this prior is shown in Fig.~\ref{fig:sed.f} (gray shaded region).
The model attempts to reproduce the spectrum (modulo the polynomial scaling), the NIRCam--MIRI photometry, and the ALMA observations, both the \oiii$\lambda88\mu$m line flux and the upper limit on the flux density of the dust continuum. The NIRSpec and NIRCam data are from \citet{Carniani:2024}, the MIRI data is from \citet{Helton:2024}. Given the redshift measurement from ALMA, we impose a strong Gaussian prior centered on $z=14.1796$ with dispersion $0.0001$.
A summary of the model parameters is reported in Table~\ref{table:2}. We infer the posterior probability distribution of our model parameters using \texttt{dynesty} \citep{speagle2020,Koposov+2022}. The 1-d marginalized posteriors are also listed in Table~\ref{table:2}. Fig.~\ref{fig:sed} reports the main 1-d and 2-d posterior probabilities (triangle diagram; Fig.~\ref{fig:sed.a}), the galaxy's SED (panel~\subref{fig:sed.b}), the normalized residuals (panel~\subref{fig:sed.c}), and the SFH (panel~\subref{fig:sed.d}).
When fitting spectroscopy, the signal-to-noise ratio of the data is much higher than for photometry, resulting in significantly smaller uncertainties. The result is that our posterior probabilities are dominated by systematic uncertainties. To estimate their magnitude, we run \textsc{prospector} five times using the same data but different priors.

With this setup, we find fiducial (magnification-corrected) $\log\,\mstar/\msun =\prospmstar({\rm ^{+0.4}_{-0.1}~systematics})$ and a $\log \mathrm{SFR} /(\msun~\peryr) = \prospsfr (\pm0.1)$. We report the star-formation rate averaged over the previous $10$ Myr.
To match the redshift of \oiii$\lambda88\mu$m with that of the Ly$\alpha$ drop, the model also requires a DLA, with $\log N_\mathrm{H\,\textsc{i}}/\mathrm{cm^{-2}} = \prospDLA (\pm 0.01)$; this value is in agreement with the fiducial DLA model which gives a value of 22.23 (Section~\ref{sec:dla}). We rule out a major role of wavelength-calibration issues between ALMA and NIRSpec, because the redshift from ALMA matches the 3-$\sigma$ tentative detection of \ciii from NIRSpec itself. The results are in agreement with the power-law fits reported in Section~\ref{sec:dla}.
We find a metallicity of 0.17~solar, somewhat higher than previous results \citep{Carniani:2024,Helton:2024}; this could be driven by the \oiii\,88$\mu$m detection.
Our setup finds a lower dust attenuation value and escape fraction than \citet{Carniani:2024}, however, these two parameters are unsurprisingly degenerate.
In our fiducial model, the model preference for a low dust attenuation requires a strong nebular continuum to explain the relatively flat UV slope reported by \citet{Carniani:2024}; this is evident from the presence of a Balmer jump in the posterior model (Fig.~\ref{fig:sed.b}). At the same time, the bright flux measured by MIRI requires a moderate contribution from the stellar continuum in addition to \hb and \oiii$\lambda\lambda4959,5007$; this stellar continuum displays a moderate Balmer break ($\approx1.4$, when measured without the nebular continuum). %The model finds a \oiii$\lambda5007$/\hb flux ratio of \prospothb (bottom right histogram in Fig.~\ref{fig:sed.a}), somewhat in tension with the values reported in \citet[][which are about two times lower]{Helton:2024}.

The ALMA line flux predicted by the model is degenerate with stellar mass, metallicity, SFR and ionization parameter; these strong degeneracies highlight the constraining power the joint use of ALMA and JWST.

The SFH follows the rising shape of the prior probability, although we note that the prior width is significantly wider than the posterior probability (gray vs red shaded regions in Fig.~\ref{fig:sed.f}). We stress that the prior range shown in the SFH is the scale parameter of the Student's $t$ distribution; the prior standard deviation is larger. The key constraints from the prior come at the earliest epochs, when the prior suppresses the earliest and least constrained stages of star formation. There is some indication of the SFH `stalling' at recent times, with the SFR being roughly constant in the last 30~Myr. This goes against the rising expectation from the prior, which would give a certain degree of confidence to this stalling behavior. However, we also find that different assumptions on the dust attenuation and prior on the SFH do impact the exact shape of the recovered SFH.

\begin{table*}
    \begin{center}
    \caption{Summary of the parameters, prior probabilities, and posterior probabilities of the fiducial \prospector SED model (see also Fig.~\ref{fig:sed.a}).
    }\label{table:2}
    \setlength{\tabcolsep}{4pt}
    \begin{tabular}{>{\tiny}ll>{\tiny}c>{\small}l>{\tiny}lc}
  \hline
   & Parameter & {\normalsize Free} & {\normalsize Description} & {\normalsize Prior} & Posterior \\
   & (1)       & (2)  & (3)         & (4)   & (5) \\
   \hline
   \multirow{11}{*}{\rotatebox[origin=c]{90}{Star-forming component}}
   & $z_\mathrm{obs}$ & Y & redshift & $\mathcal{N}(z_\mathrm{spec}, 0.001)$ & $14.178^{+0.001}_{-0.001}$ \\
   & $\log \mstar [\msun\,\mu^{-1}]^\dagger$ & Y & total stellar mass formed & $\mathcal{U}(6, 10)$ & \prospmstar $(^{+0.4}_{-0.1})$\\
   & $\log Z [\zsun]$ & Y & stellar and gas metallicity & $\mathcal{U}(-2, 0)$ & \prosplogz $(\pm 0.03)$ \\
   & $\log \mathrm{SFR}$ ratios & Y & ratio of the $\log \mathrm{SFR}$ of non-parametric SFH & $\mathcal{T}(R(z)^\ddag, 1, 2)$ & --- \\
   & $\sigma_\star \; [\kms]$ & Y & stellar intrinsic velocity dispersion & $\mathcal{U}(0, 300)$ & $190^{+70}_{-70}$ \\
   & $n$ & Y & power-law modifier of the dust curve \citepalias[][eq.~5]{Tacchella:2022a} & $\mathcal{G}(0,0.1;-1.0,0.2)$ & $0.06^{+0.05}_{-0.04}$ \\
   & $\tau_V$ & Y & optical depth of the diffuse dust \citepalias[][eq.~5]{Tacchella:2022a} & $\mathcal{U}(0,2)$ & \prosptauv \\
   & $\mu_\mathrm{d}$ & Y & ratio of optical depths of the birth clouds and $\tau_V$ & $\mathcal{G}(1,0.1;0,2)$ & $0.94^{+0.06}_{-0.07}$ \\
   & $\sigma_\mathrm{gas} \; [\kms]$ & Y & intrinsic velocity dispersion of the star-forming gas & $\mathcal{U}(0,1500)$ & $1,310^{+90}_{-100}$ \\
   %& $\log Z_\mathrm{gas} [\Zsun]$ & Y & metallicity of the star-forming gas & $\mathcal{U}(-2, 0.19)$ & $0.03^{+0.02}_{-0.02}$ \\
   & $\log U$ & Y & ionization parameter of the star-forming gas & $\mathcal{U}(-4, -1)$ & $-2.47^{+0.24}_{-0.18}$\\
   & $\fesc$  & Y & birth-cloud escape fraction of ionizing radiation & $\mathcal{U}(0, 1)$ & $0.11^{+0.09}_{-0.07}$\\
   & $\log N_{\mathrm{H}\,\textsc{i}}$  & Y & column density of neutral hydrogen & $\mathcal{U}(17, 25)$ & \prospDLA $(\pm0.1)$\\
%   \midrule
%   \multirow{6}{*}{\rotatebox[origin=c]{90}{Shock component}}
  \hline
   \multirow{3}{*}{\rotatebox[origin=c]{90}{Other}}
   & $j_\mathrm{spec}$ & Y & multiplicative noise inflation term for spectrum & $\mathcal{U}(0.5,2)$ & $1.5^{+0.3}_{-0.4}$ \\
   & $\log SFR_{10} [\msun \, \peryr \, \mu^{-1}]^\dagger$ & N & SFR averaged over the last 10~Myr & --- & \prospsfr $(\pm 0.05)$ \\
   & $\log SFR_{100} [\msun \, \peryr \, \mu^{-1}]^\dagger$ & N & SFR averaged over the last 100~Myr & --- & $0.29^{+0.14}_{-0.09} (\pm 0.1)$ \\
  & $A_{\rm v} [mag]$ & N & optical dust attenuation & --- & $0.19^{+0.07}_{-0.05} (\pm 0.05)$  \\
   & $F_{\oiii} [10^{-19}\,\mathrm{erg\,s^{-1}\,cm^{-2}}$] & N & FIR \oiii line flux & --- & $2.77^{+0.79}_{-0.78} (\pm0.05)$ \\
%   & $\oiii\lambda 5007/\mathrm{H}\beta$ & N & Emission-line ratio & --- & \prospothb \\
   & $age\; [\mathrm{Myr}]$ & N & Mass-weighted stellar age & --- & $40\pm5$\\
  \hline
  \end{tabular}
  \end{center}
(1) Parameter name and units (where applicable). (2) Only parameters marked with `Y' are optimized by \prospector; parameters marked with `N' are either tied to other parameters (see Column~4), or are calculated after the fit from the posterior distribution (in this case, Column~4 is empty). (3) Parameter description. (4) Parameter prior probability distribution; $\mathcal{N}(\mu, \sigma)$ is the normal distribution with mean $\mu$ and dispersion $\sigma$; $\mathcal{U}(a, b)$ is the uniform distribution between $a$ and $b$; $\mathcal{T}(\mu, \sigma, \nu)$ is the Student's $t$ distribution with mean $\mu$, dispersion $\sigma$ and $\nu$ degrees of freedom; $\mathcal{G}(\mu, \sigma; a, b)$ is the normal distribution with mean $\mu$ and dispersion $\sigma$, truncated between $a$ and $b$.
(5) Median and 16\textsuperscript{th}--84\textsuperscript{th} percentile range of the marginalised posterior distribution; for some nuisance parameters we do not present the posterior statistics (e.g., log SFR ratios). $^\dagger$ We corrected all extensive quantities for gravitational lensing, using $\mu=1.17$ \citep{Carniani:2024}. $^\ddag$ The non-parametric SFH is expressed by the logarithm of the SFR between adjacent time bins; we use a `rising' SFH probability prior, which at each time bin is a $\mathcal{T}$ distribution with mean log ratio $R(z)\equiv \log\,\left( SFR(z_i) / SFR(z_{i+1}) \right)$, where $z_i$ is the redshift of the $i$\textsuperscript{th} SFH bin and the $SFR(z)$ is given by \cite{Turner:2024}, their eq.~(4).
\end{table*}
% The  \Loiii/SFR L[OIII]/SFR and L[CII]/SFR
% relation of ID4590. Given the purpose of the analysis, we use the [C ii] results at the galaxy position. For
% comparison, we also present the relations from observations (left panel) and the predictions from the photoionization model with cloudy (right panel) 

\subsection{Dynamical mass}\label{sec:Mdyn}

Given that we spectroscopically resolve the emission line in the ALMA observations, we are now able to constrain the dynamical mass (M$_{\rm dyn}$), assuming that the system is close to virialization, and can compare it to the stellar mass from the \texttt{prospector} SED fitting. 
% To estimate the dynamical mass, we use the size of the galaxy and its mass profile (probed by the S{\'e}rsic index) inferred by \cite{Carniani:2024}.
We estimated the dynamical mass following the approach described by \citet{Ubler:2023} using the equation:
\begin{equation}
    M_{\rm dyn} = K(n) K(q) \frac{\sigma^{2}R_{e}}{G}
\end{equation}
where K(n) =  8.87 -- 0.831n + 0.0241n$^2$ and with S\'{e}rsic index n, following \citet{Cappellari2006}, K(q) = [0.87 +
0.38e$^{-3.71(1-q)}]$$^{2}$, with axis ratio q following \citet{van_der_Wel2022}, $\sigma$ is the integrated stellar velocity dispersion, and R$_{e}$ is the effective radius.
We remark that the above approach was calibrated for stellar kinematics of massive galaxies at $z=0.8$ ($\mstar= 10^9\text{--}10^{11}~\msun$). However,  a range of possible calibrations \citep[e.g.,][]{Wisnioski:2018,Cappellari:2013} provides similar answers, within 0.3~dex \citetext{see e.g. \citealp{Marconcini+2024}}. 
We use our measured value of $\sigma = {\rm FHWM} / 2.355 = 43.5^{+12.3}_{-9.2}$~\kms from \oiii 88\,$\mu$m, to which we apply a correction of $\Delta \log (\sigma/(\kms))$ = +0.1 following \cite{Ubler:2023}, since galaxies with low integrated ionized gas velocity dispersion tend to have higher integrated stellar velocity dispersion \citep{Bezanson:2018}.
We adopt a R$_{e}$ value of 260$\pm$20 pc from \citet{Carniani:2024}. For JADES-GS-z14-0, \citet{Helton:2024} reports $n = 0.877 \pm 0.027$ for the S\'{e}rsic index and $q =  0.425 \pm 0.008$ for the axis ratio. However, since it is unclear whether the stellar emission in the rest-frame UV has the same morphological properties as the nebular emission in the rest-frame far-infrared, we adopt a range of values for the S\'{e}rsic index and axis ratio (n=0.8--2 and q=0.3--1).
We thus estimate a log(M$_{\rm dyn}/\msun$) = 9.0$\pm0.2$.

This value of M$_{\rm dyn}$ is five times larger than the stellar mass from our SED fitting \citetext{$\log \mstar/\msun = \prospmstar$, which assumes a \citealp{Chabrier:2003} IMF}. Assuming a disc-like geometry with a mean inclination of $i=60$\textdegree, this would imply a fraction of gas mass and dark matter in this galaxy of $\sim 80\%$. However, we stress that this gas mass fraction is an upper limit. If we adopt the stellar mass estimates inferred by \cite{Helton:2024}, who explored different star-formation histories, we infer a gas mass fraction lower than 50\%-60\%. Alternatively, if the inclination was very small (close to face on) or the \oiii extent is significantly broader than that of UV continuum emission, the line broadening could underestimate the dynamical mass. Observations with higher spatial resolution would be required to address this scenario.

\section{A non-zero escape fraction?}\label{sec:escape}

The luminosity of the \oiii far-infrared line of \gs indicates a gas-phase metallicity of about 0.2~$Z_{\odot}$, which is similar to that measured in other luminous galaxies such as GN-z11 \citep{Bunker:2023, Maiolino:2023} and GHZ2 \citep{Castellano:2024, Zavala:2024, Calabro:2024}. Although these three galaxies have similar UV luminosities and metal enrichment conditions, most of the UV lines of \gs are undetected in the deep NIRSpec observations. The upper limits on the rest-frame equivalent widths of \gs are up to seven times lower with respect to those measured in GN-z11 and GHZ2. 

\begin{figure}
    \centering
    \includegraphics[width=0.4\textwidth]{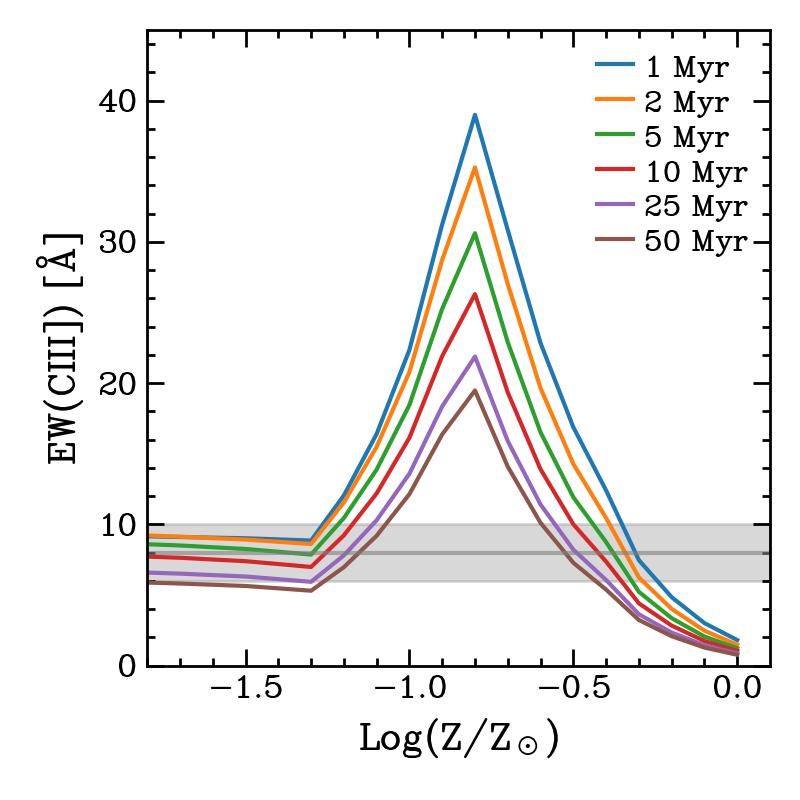}
    \caption{ EWs of \ciii as a function of gas-phase metallicity. The blue, orange, green, red, purple, and brown curves denote the metallicity dependencies for exponential SFH (see Appendix~\ref{app:balmer_break}) with stellar ages of  1, 2, 5, 10, 25, and 50 Myr, respectively. The gray line and shaded region represent the \ciii EW measurement of \gs. The observed \ciii EW is typical of galaxies with $Z>0.3{\rm Z_\odot}$ or $Z<0.05{\rm Z_\odot}$}
    \label{fig:EWciii}
\end{figure}

A non-zero escape fraction of ionizing photons might explain the low intensity of UV emission lines in \gs. This scenario is also supported by the SED fitting of the combined ALMA and JWST datasets that report an escape fraction of at least 10\%. Figure~\ref{fig:EWciii} shows the equivalent of \ciii as a function of gas-phase metallicity from \textsc{prospector} using exponential star-formation histories with stellar ages from 1 Myr to 50 Myr (see details in Appendix~\ref{app:balmer_break}. Independently from the stellar age, the EW of the UV line increases with decreasing metallicity down to 0.15~${\rm Z_\odot}$ and then decreases because carbon abundance decreases. To reproduce the strength of \ciii of \gs, the gas phase metallicity should be either lower than 0.05~${\rm Z_\odot}$ or higher than 0.3~${\rm Z_\odot}$. The latter was considered less probable by  \cite{Carniani:2024} and \cite{Helton:2024} because of the intensity of the MIRI flux and the stellar age of the galaxy. However, the new detection of the far-infrared oxygen line combined with the current SFR of the galaxy puts a stringent limit of 0.1-0.3~${\rm Z_\odot}$ to the gas-phase metallicity, which corresponds to the metallicity range excluded initially by the analysis based solely on the UV lines. With the new constraints from ALMA, \textsc{prospector} thus requires an escape fraction of $\sim10\%$ to match the two apparently discordant observables of a relatively high \oiii flux but a relatively low EW of \ciii. We stress that -- like all SED fitting, we are assuming here a single-zone model, with a reference electron density and solar abundances \citep{Byler2017}.

The origin of such a high escape fraction can be the result of fast galactic outflows that expel gas to a large scale and carve the ISM of the galaxy. For example, \cite{Ferrara:2024} predict that \gs may have recently experienced a starburst phase and radiation-driven outflows have cleared the galaxy from dust and gas. Current JWST and ALMA data do not have the sensitivity to identify the presence of ``wings'' in the emission line profiles tracing outflowing gas. However, the analysis of the dynamical mass suggests that the potential well is dominated by the stellar population and a large fraction of the gas may have been removed from the galaxy. If the observed $\log(N_{\rm HI}/\mathrm{cm}^{-2})=22.23$ DLA described in Section \ref{sec:dla} was a result of a potential galactic outflow, this would help explain its presence while also allowing for a high escape fraction of ionizing radiation. Another possible explanation is the ionizing photons escape along a different direction than that of DLA systems and are not absorbed (see \citealt{Tacchella:2024} for the $z=5.94$ galaxy GS9422).

Intriguingly, the implied $10\%$ escape fraction in a source with the luminosity of JADES-GS-z14-0 would imply the rapid formation of a local ionised `bubble' allowing Ly$\alpha$ emission to escape through the intergalactic medium, as required for JADES-GS-z13-1-LA observed to have a bright Ly$\alpha$ line at $z = 13$ \citep{Witstok:2024b}.

\section{Carbon-to-oxygen abundance ratio}\label{sec:co}

Leveraging the redshift delivered by the detection of \oiii$88\mu$m in the ALMA observations, we can confirm the tentative detection of \ciii previously reported by \cite{Carniani:2024} in the JWST/NIRSpec spectrum at $2.89\mu$m.
This allows us, under a number of fiducial assumptions discussed below, to derive the carbon-over-oxygen abundance (C/O) for \gs, yielding the most distant measurement of such abundance ratio to date.

To derive C/O, we first corrected the measured \ciii flux for nebular attenuation assuming A$_\text{V}$=0.25 as inferred from our SED modelling\footnote{We used A$_V\equiv \tau_V \cdot (1 + \mu_\mathrm{d}) \cdot 1.086$.}
We then derive the C$^{++}$/O$^{++}$ abundance from the \ciii/\oiii$88\mu$m line ratio adopting \textsc{pyneb} \citep{luridiana_pyneb_2015}, assuming an electron temperature of T$_{e}$=$15,000$~K and an electron density of n$_{e}$=$100$~cm$^{-3}$. The assumed T$_{e}$ value is consistent with the best-fit metallicity and ionisation parameter inferred from the SED modelling, and we assume to reside in the low-density regime as suggested by the simultaneous modelling of the \oiii$88\mu$m/\oiii5007 and \oiii5007/\hb ratios (see Figure~\ref{fig:o33o3hb}).
We then correct the C$^{++}$/O$^{++}$ abundance to the total C/O by applying an ionisation correction factor (ICF) derived from the models presented in \cite{Berg2024}, under the assumption of log(U)=$-2.4$ and Z=0.2~Z$_{\odot}$ (i.e., the best-fit values from the SED fitting).
The total C/O value is therefore log(C/O)$=-0.72\pm0.13$, corresponding to [C/O]=log(C/O)-log(C/O)$_{\odot}$ = --0.36, where the formal uncertainties are computed by randomly varying the measured fluxes by their errors and repeating the calculation 100 times, taking the standard deviation of the resulting C/O distribution.

\begin{figure*}
    \centering
    \includegraphics[width=0.43\textwidth]{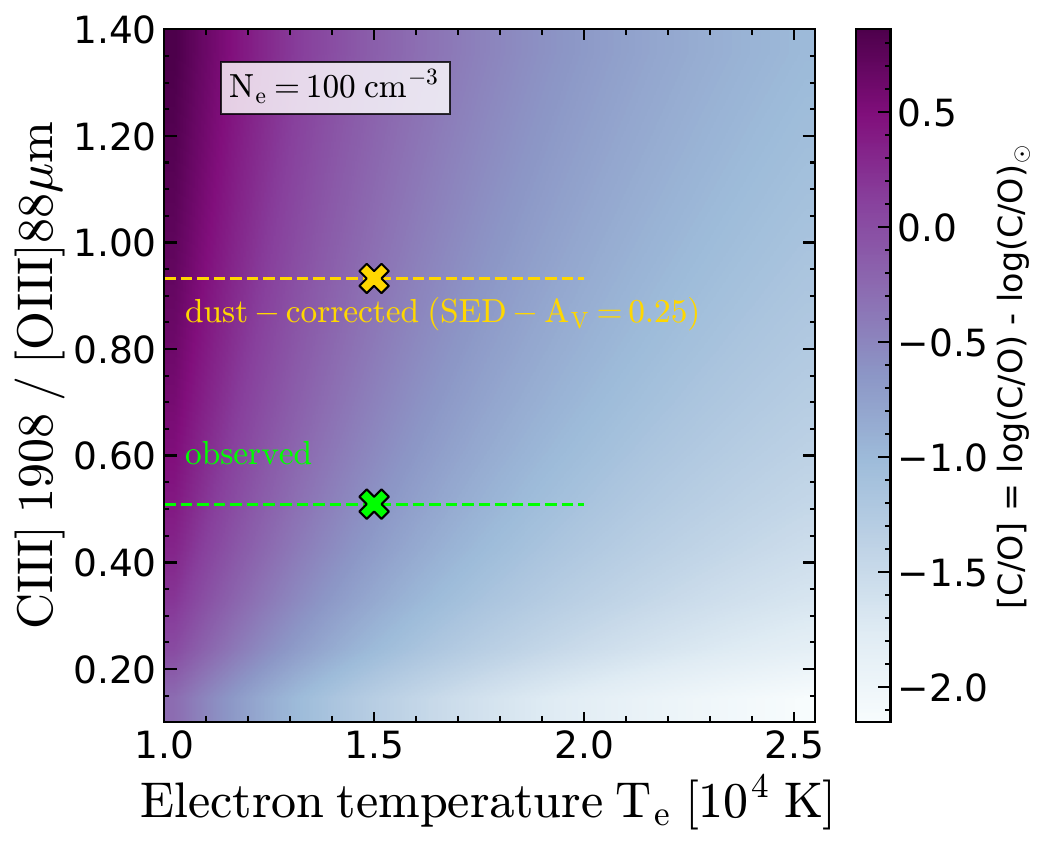}
    \includegraphics[width=0.5\textwidth]{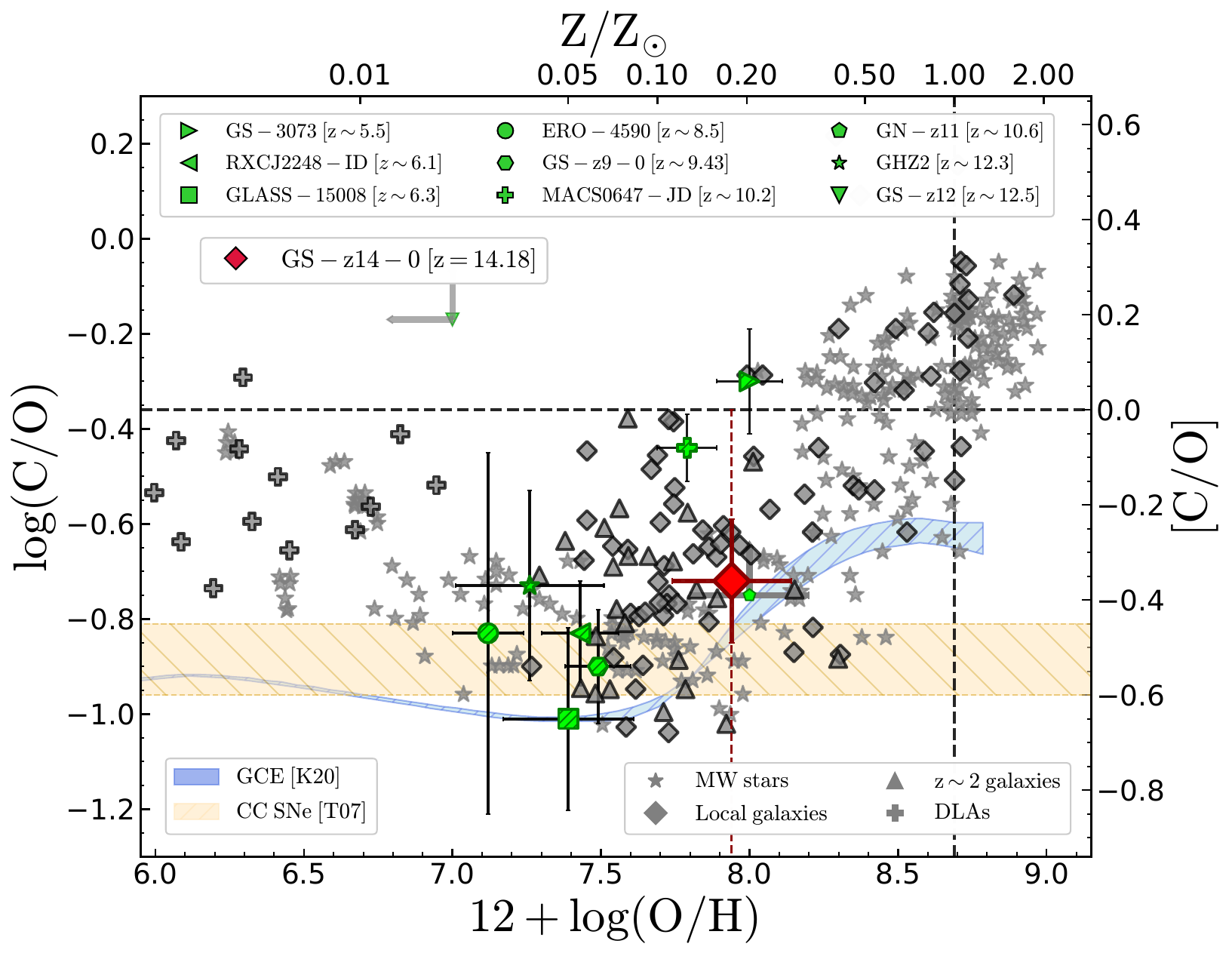}
    \caption{Carbon-over-oxygen abundance for \gs. The left-hand panel shows how the final inferred C/O value changes as a function of the \ciii/\oiii$88\mu$m line ratio and of the electron temperature, T$_e$, under our fiducial assumption of N$_{e}$=$100$~cm$^{-3}$. The observed and dust-corrected \ciii/\oiii$88\mu$m ratios for \gs are marked in green and yellow, respectively. In the right-hand panel, we report our fiducial C/O measurement for \gs on the C/O vs O/H diagram. The oxygen abundance is inferred from the best-fit \textsc{prospector} metallicity reported in Table~\ref{table:2}. The C/O value measured in \gs is consistent with pure enrichment from core-collapse Supernovae. We also report for comparison a sample of C/O measurements from \textsc{JWST} compiled from the literature, namely GS-z12 (z = 12.5; \citealt{DEugenio:2023}), GHZ2 (z = 12.34; \citealt{Castellano:2024}), GN-z11 (z = 10.6; \citealt{Cameron:2023}), MACS0647-JD (z = 10.2; \citealt{Hsiao:2024}), GS-z9-0 (z = 9.4; \citealt{Curti24}), ERO-4590 (z = 8.5; \citealt{Arellano-Cordova22}), RXCJ2248-ID (z = 6.11; \citealt{Topping:2024}), GLASS-150008 (z = 6.23; \citealt{Jones23}), GS-3073 (z=5.5; \citealt{Ji24}). The C/O vs O/H pattern predicted by Galactic Chemical Evolution models of \cite{kobayashi_origin_2020} is shown in blue, whereas the C/O range allowed by the CC-SNe yields from \cite{tominaga_2007} is marked by the golden region.
    }
    \label{fig:CO}
\end{figure*}

Nonetheless, we note that variations in our assumptions on both T$_{e}$ and n$_{e}$, as  well as uncertainties associated with the dust attenuation correction, can have a non-negligible impact onto the final C/O.
In the left panel of Figure~\ref{fig:CO}, we show how the final inferred C/O value would change as a function of the \ciii/\oiii$88\mu$m line ratio (comparing in particular the observed and dust-corrected value based on the SED-inferred A$_\text{V}$) and of the electron temperature, under our fiducial assumption of n$_{e}$=$100$~cm$^{-3}$.
For instance, at fixed \ciii/\oiii$88\mu$m ratio and n$_{e}$, varying T$_{e}$ between $15,000$~K (our fiducial value) and $20,000$K lowers the inferred C/O by $\sim0.45$~dex.
On the other hand, assuming n$_{e}$=300~cm$^{-3}$ instead of 100~cm$^{-3}$ (at fixed line ratio and T$_{e}$) would lower the C/O by $\sim0.18$~dex.
We therefore estimate the amount of systematic uncertainty on our C/O derivation associated with different assumptions on density and temperature by computing the C/O values over a grid in n$_{e}$ and T$_{e}$ spanning n$_{e}$ $\in$ [100,300]~cm$^{-3}$ and T$_{e}$ $\in$ [10,000, 20,000]~K, yielding a final log(C/O)$=-0.72\pm0.13~(\substack{+0.3\\-0.6}$~systematics). 

In the right-panel of Figure~\ref{fig:CO}, we finally plot \gs on the C/O vs O/H diagram as the red diamond marker. The solid errorbars report the formal statistical uncertainty on C/O from the flux measurement errors ($0.13$~dex), whereas the dashed errorbars encompass the systematic uncertainties ($\substack{+0.3\\-0.6}$) from our assumptions on n$_{e}$ and T$_{e}$ as discussed above.
The oxygen abundance is inferred from the best-fit \textsc{prospector} metallicity.
For comparison, a compilation of C/O measurement at high-z from \textsc{JWST} spectroscopy is also shown (see the legend and caption of Figure~\ref{fig:CO} for details).

The fiducial C/O value measured in \gs appears in line with expectations given its moderate metallicity, and it is consistent within the uncertainties with predictions from galactic chemical evolution models of the solar neighborhood \citep{kobayashi_origin_2020}, as well as with the yields from pure core-collapse Supernovae (CC-SNe) enrichment \citep{tominaga_2007}, in agreement with a rapid and recent history of mass assembly. 
Nonetheless, we note that even slightly higher C/O values than predicted by pure CC-SNe enrichment, as possibly shown by taking our C/O measurement in \gs at face value, might be indicative of a lower effective oxygen yield, as possibly driven by SNe-driven outflows preferentially expelling oxygen from the ISM following an intense phase of star-formation (as also discussed in Sect.~\ref{sec:escape}), whereas any contribution to Carbon enrichment from evolved, low-mass stars is unlikely given the age of the system.

As a caveat, we reiterate that our C/O measurement assumes that \ciii and \oiii$88\mu$m trace regions with similar densities. However, the carbon line can originate from regions with densities exceeding the critical density of \oiii$88\mu$m, potentially affecting the C/O abundance estimate. In the presence of multiple density components, the C/O ratio can only be reliably measured using emission lines with comparable critical densities. In this context, deeper observations of JADES-GS-z14-0 with NIRSpec/JWST have the potential to detect O\,{\sc iii}]$\lambda\lambda1661,1666$, which, when combined with \ciii, would enable a more precise determination of the carbon-to-oxygen ratio.

\section{Conclusions}

We have presented the analysis of the new DDT ALMA program targeting the \oiii\ 88\,$\mu$m emission in the most distant and luminous galaxy known so far,  \gs. 
We have detected the FIR line with a level of significance of 6.67$\sigma$ at a frequency of 223.524~GHz. We have thus determined the spectroscopic redshifts of $z_{\rm [OIII]} = 14.1796 \pm 0.0007$, in excellent agreement with that from the tentative detection of \ciii ($z_{\rm CIII]} = 14.178 \pm 0.013$), and consistent with the Lyman-break redshift ($z_{\rm Ly\alpha} = 14.32^{+0.08}_{-0.20}$) when accounting for a DLA ($\log N_{\mathrm{H}\,\textsc{i}}/\mathrm{cm}^{-2}= \prospDLA$).

The measured luminosity of the line ($L_{\rm [OIII]}/\lsun = 8.3\pm 0.1$) is consistent with the $L_{\rm  [OIII]}-SFR$  relation of local dwarf galaxies. The detection of the FIR line indicates a gas-phase metallicity $Z>0.1~Z_\odot$ and thus suggests a rapid metal enrichment during the earliest phases of galaxy formation. The modest metal enrichment measured in this work is somewhat unexpected, given the lack of most of the rest-frame UV emission lines in the deep NIRSpec spectrum \citep{Carniani:2024}.  However, the discrepancy between the two observations can be mitigated by assuming a non-zero escape fraction ($\sim10\%$), which would explain the low intensity of UV emission lines despite the modest gas-phase metallicity measured in \gs. 

We have combined the MIRI flux at 7.7~$\mu$m and the ALMA detection to constrain the interstellar medium condition by using predictions from  \textsc{Cloudy} models. The excess of MIRI flux, likely associated to the nebular optical lines of oxygen and hydrogen, and the FIR $L_{\rm  [OIII]}$ set a stringent upper limit on the electron density of $n_{\rm e}<700~{\rm cm^{-3}}$ at $Z = 0.2~Z_\odot$,
which is lower than those previously found in several [O\textsc{ii}]-emitting sources at $z>9$ \citep[e.g.,][]{abdurrouf_2024ne,Marconcini+2024,Zavala:2024}, 
and also lower compared to the average density ($n_{\rm e}\sim1000~{\rm cm^{-3}}$) expected at $z>9$ from the extrapolation of the [O\textsc{ii}]-derived $n_{\rm e}$ vs. $z$ relation by \citet{isobe_nez2023}. 
One possible physical reason is that galactic outflows may have expelled gas on large scales and reduced the gas density within the galaxy.

To constrain the properties of the stellar population and interstellar medium we have performed the SED fitting with \textsc{prospector} to the JWST and ALMA data. The inferred star-formation rate and stellar mass are in agreement with those inferred by \cite{Carniani:2024} and \cite{Helton:2024}. We note that  \textsc{prospector} results indicate a gas-phase metallicity 17\% solar, but it requires an escape fraction of 0.1 to match the NIRSpec spectrum and, therefore, the weakness of the UV emission line.
Combining the \oiii88$\mu$m detection in ALMA with the $3.5\sigma$ \ciii emission detected by JWST/NIRSpec, we also derive a carbon-over-oxygen abundance log(C/O)$=-0.72\pm0.13~(\substack{+0.3\\-0.6}$), in agreement with chemical evolution models of the solar neighborhood given the moderate enrichment of the system.

We have spectroscopically resolved the \oiii\  allowing us to measure the dynamical mass of this galaxy of log(M$_{\rm dyn}$) = 9.0$\pm0.20$. We find this measurement consistent with the stellar mass estimated from \textsc{prospector}, implying a gas fraction lower than 80\%. However, we note that adopting different priors on the star formation history in the SED fitting process \cite{Helton:2024} results in an inferred gas mass fraction below 50\%–60\%. We speculate that galactic outflows may have carved the interstellar medium, hence reducing the gas reservoir and resulting in a non-zero escape fraction as expected by \textsc{prospector} results.

This study demonstrates that multi-wavelength observations are necessary to constrain the properties of galaxies at the highest redshifts. However, the detection of only one emission line is not sufficient to put stringent constraints on the interstellar medium condition. MIRI spectroscopic observations and deep NIRSpec data would allow us to investigate directly the luminosity of a range of optical and UV emission lines. Combined with ALMA, these lines would provide new insight on the metal enrichment conditions of the galaxy. Deeper and higher frequency ALMA observations capable of probing the dust continuum would also be invaluable, by providing crucial information on the dust content and attenuation.

%----------------------------------------------------------------- 
   % \begin{figure}
   % \centering
   % \includegraphics[width=\linewidth]{figures/temden_gsz140.pdf}
   %    \caption{Electron densities and temperatures constrained by the flux ratio of \oiii$5007\AA$/\oiii$88\mu$m.
   %    Model grids are computed using \textsc{PyNeb}, assuming atomic data from \textsc{CHIANTI} and Tayal \& Zatsarinny (2017).
   %    }
   %       \label{fig:temden}
   % \end{figure}
%-----------------------------------------------------------------

\begin{acknowledgements}
      This paper makes use of ALMA data: 2023.A.00037.S. ALMA is a partnership of ESO (representing its member states), NSF (USA) and NINS (Japan), together with NRC (Canada) and NSC and ASIAA (Taiwan), in cooperation with the Republic of Chile. The Joint ALMA Observatory is operated by ESO, AUI/NRAO and NAOJ. 
      S.C, EP and GV acknowledge support from the European Union (ERC, WINGS,101040227). 
      FDE, XJ, JS, RM and JW acknowledge support by the Science and Technology Facilities Council (STFC), ERC Advanced Grant 695671 ``QUENCH" and the UKRI Frontier Research grant RISEandFALL. RM also acknowledges funding from UKRI Frontier Research grant RISEandFALL and a research professorship from the Royal Society.
      AJB and AS acknowledge funding from the "FirstGalaxies" Advanced Grant from the European Research Council (ERC) under the European Union’s Horizon 2020 research and innovation programme (Grant agreement No. 789056).
      H{\"U} gratefully acknowledges support by the Isaac Newton Trust and by the Kavli Foundation through a Newton-Kavli Junior Fellowship.
      The authors acknowledge use of the lux supercomputer at UC Santa Cruz, funded by NSF MRI grant AST 1828315.
      CNAW and DJE are supported as a Simons Investigator and by JWST/NIRCam contract to the University of Arizona, NAS5-02015.
      The Cosmic Dawn Center (DAWN) is funded by the Danish National Research Foundation under grant DNRF140.
      BER and BDJ acknowledge support by JWST/NIRCam contract to the University of Arizona, NAS5-02015 and BDJ is supported by JWST Program 3215. 
      The research of CCW is supported by NOIRLab, which is managed by the Association of Universities for Research in Astronomy (AURA) under a cooperative agreement with the National Science Foundation.

\end{acknowledgements}

  \bibliographystyle{aa} % style aa.bst
  \bibliography{biblio} % your references Yourfile.bib

\newpage

\begin{appendix}
\section{Fidelity of ALMA \texorpdfstring{\oiii}{O3}\ line detection}
\label{apd:stats} 

To assess the fidelity of ALMA \oiii\,88\,$\mu$m line detection at 223.524\,GHz, we have blindly searched for positive and negative peaks in reduced ALMA spectral cube.
Using the natural cube, we first measure noise spectrum from each channel (median: 0.15\,mJy\,beam$^{-1}$ per 10 km\,s$^{-1}$ velocity bin).
We then bin the spectral cube with various velocity sizes of 30, 50, ..., 170\,km\,s$^{-1}$, respectively, and search for positive and negative peaks in the collapsed spectral cubes over the central $25^{\prime\prime}\times25^{\prime\prime}$ region.
The S/N of detected peaks is computed by the peak brightness divided by the noise spectrum collapsed over the same velocity bin.

\begin{figure}[h]
\centering
\includegraphics[width=0.85\linewidth]{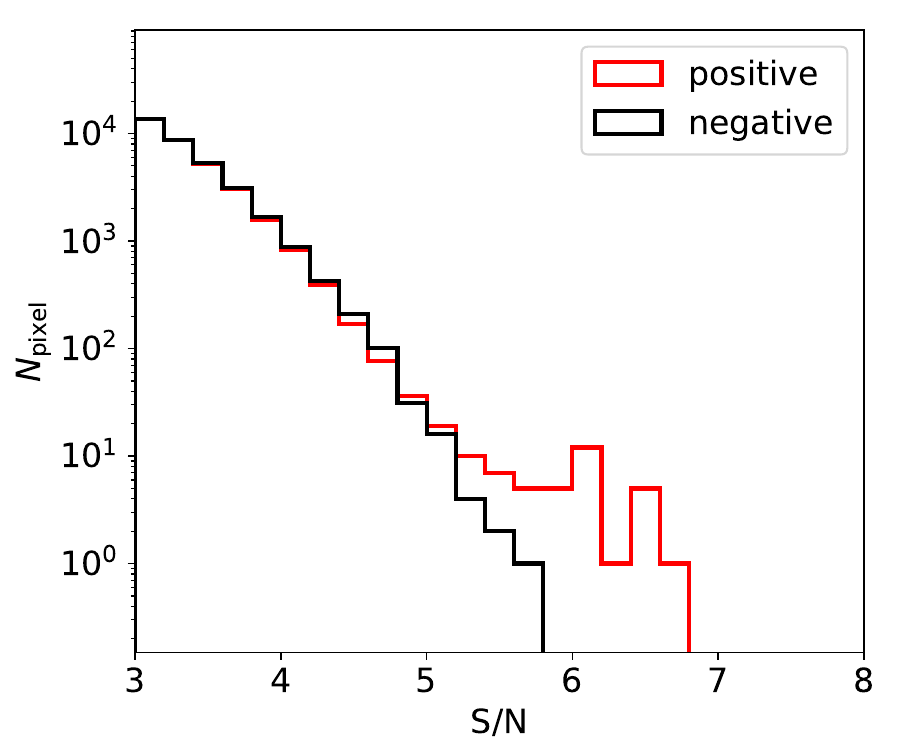}
\caption{Statistics of positive and negative peaks from natural-weighted spectral cubes, collapsed with various velocity bin sizes at 30--170\,km\,s$^{-1}$.
The maximum S/N\,=\,6.67 is seen at the centroid of \gs\ at 223.524\,GHz with a velocity bin size of 130\,km\,s$^{-1}$, for which we interpret as an \oiii\,88\,$\mu$m line detection at $z=14.1796$.
This is much more significant than the strongest negative peak seen in the spectral cube ($-5.7\sigma$).
Therefore, we conclude the fidelity of ALMA line detection.
}
\label{fig:peak_stats}
\end{figure}

Figure~\ref{fig:peak_stats} summarizes the statistics of positive and negative peaks throughout the collapsed spectral cubes. 
% Note that same peaks could be repeated through such an experiment.
The maximum S/N of the collapsed spectral cubes is seen at the centroid of \gs\ with a velocity bin size of 130\,km\,s$^{-1}$ at a central frequency of 223.524\,GHz, for which we interpret as an \oiii\,88$\mu$m emission line detection at $z=14.1796\pm0.0007$.
We also examine the noise spectrum and conclude no additional telluric absorption or contamination is present at this frequency.
This line detection is much more significant than the strongest negative peak seen within the spectral cube, which is at S/N\,$=$\,--5.73 with velocity bin size of 50\,km\,s$^{-1}$.
We have also experimented the Briggs-weighted (robust=0.5) spectral cube and the conclusion is similar, although the line S/N is slightly reduced to 6.22.
The reduced S/N in Briggs-weighted cube with smaller synthesized beam size may indicate that the \oiii\,88\,$\mu$m\ line of \gs\ is spatially extended.
Although this is consistent with the spatially extended nature of \gs\ in the rest-frame UV, the current line S/N is not high enough to ensure a robust size measurement of \oiii\,88\,$\mu$m\ emission.
Nevertheless, we conclude the fidelity of the ALMA emission line presented in this work.

\section{Balmer break}
\label{app:balmer_break}
\cite{Helton:2024} determine that at least one-third of the flux detected with the MIRI filter 770W comes from the rest-frame optical emission lines \hb\ and optical \oiii doublet. However, determining the intensity of these optical nebular lines requires follow-up MIRI observations in Low-Resolution Spectroscopy  mode to disentangle the continuum emission from the line emission component. Indeed. the excess flux of 27.5 nJy in F770W with respect to F444W \citep{Helton:2024} is not a direct measurement of the \hb+\oiii line intensity as the flux level of the underlining continuum emission at 7.7$\mu$m might be different from that measured in the  F444W filter. The Balmer break, which is the ratio between the continuum emission at $\lambda_{\rm rest} = 4200~\AA$ and that at $\lambda_{\rm rest} = 3500~\AA$ depends mainly on the stellar age. \cite{Wilkins:2023} shows that the luminosity ratio $\log_{10}(L_{\rm 4200}/L_{\rm 3500})$ is as large as -0.3 for a stellar population with a stellar age of 1 Myr and reaches a value of about 0.5 for a galaxy with an average stellar age of 1 Gyr. The authors also emphasized that the strength of the Balmer break is also influenced by the shape of the star-formation history.

\begin{figure}
\centering
\includegraphics[width=0.9\linewidth]{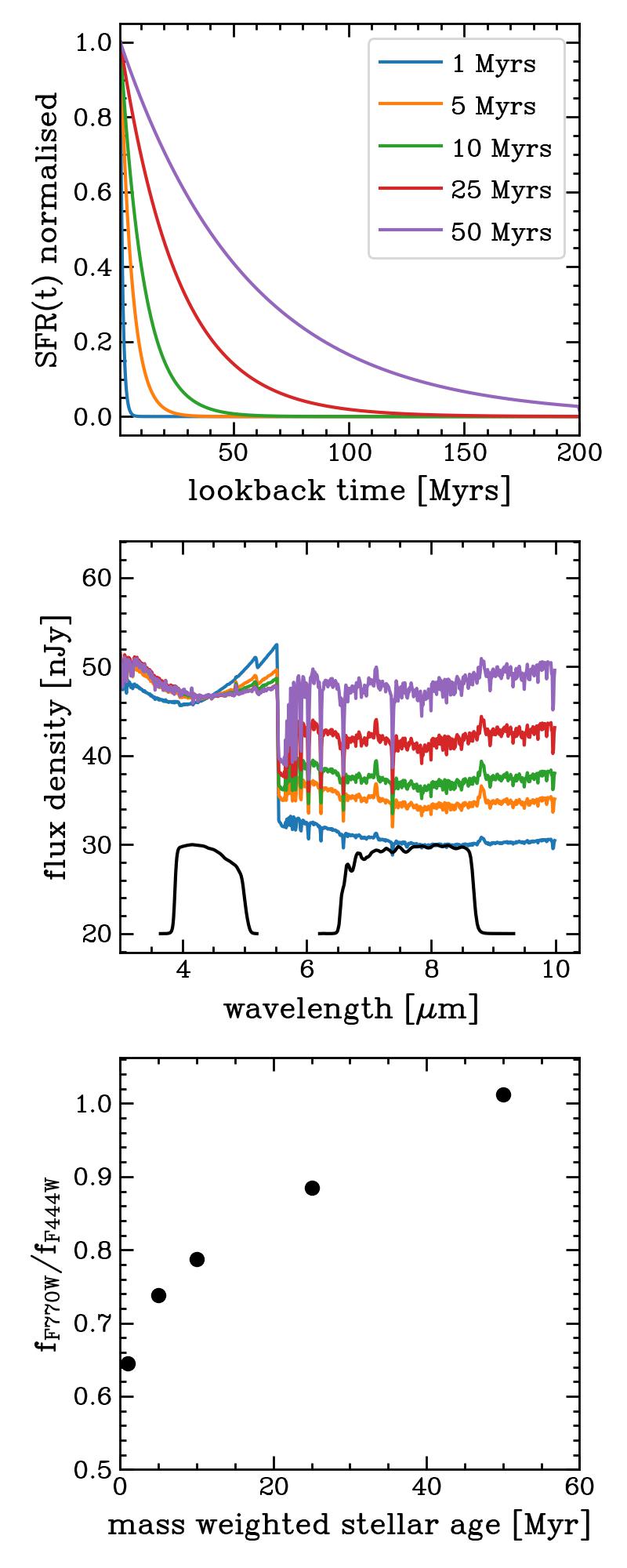}
\caption{Top: star-formation histories (SFHs) with different mass-weighted stellar age and of the form $SFR(t)\propto\exp^{-t/\tau}$. The line colors correspond to the stellar age reported in the legends Middle: SED models after removing emission lines for each SFHs and color-coded depending on the stellar age reported in the top panel. All SEDs are normalized to the flux density level of 46.9 nJy in the NIRCam F444W filter.  The black lines show the transmission curves for F444W (left) and F770W (right) JWST filters. Bottom: flux density ratio between the F770W and F444W filters as a function of the mass-weighted stellar age.} 
\label{fig:model_balmer}
\end{figure}

 To constrain the intensity of the optical \hb+\oiii line of \gs, we have evaluated the dependence of the continuum emission at 7.7~$\mu$m on the star-formation history. In particular, we have assumed five different exponential star-formation histories of the form $SFR(t)\propto\exp^{-t/\tau}$ where $t$ corresponds to the time before the observations (i.e, lookback time) and $\tau$ has been selected to return mass-weighted stellar ages of 1, 5, 10, 25, and 50 Myrs (Fig.~\ref{fig:model_balmer}). We have generated a mock galaxy spectrum for each star-formation history by using BPASS models \cite{eldridge_bpass17,stanway_bpass18,byrne_bpass22} for the stellar population and the nebular line and continuum component with CLOUDY. 
We have then normalized the spectra to the photometric value $46.9 nJy$, calculated within the wavelength range covered by the NIRCam F444W filter \cite{Robertson:2023, Helton:2024, Carniani:2024}. Finally, we have removed the emission line from the spectra and estimated the flux density in the MIRI filter F770W. The bottom panel of Fig.~\ref{fig:model_balmer} illustrates the flux density in the MIRI filter with respect to the NIRCam flux density at 4.5 $\mu$m. We note that the two flux densities are comparable when the mass-weighted stellar age of the galaxy is larger than 40-50 Myrs. In the extreme case where the mass-weighted stellar ages are of the order of 1 Myr, the continuum flux density level in the MIRI filter is about 0.65 times lower than that measured in the NIRCam wavelength range. In this case, the equivalent width of the nebular optical lines can be higher than that estimated assuming that the continuum flux level at 7.7~$\mu$m is as large as the continuum at 4.5~$\mu$m.

\end{appendix}

\end{document}